%
%
\input mtexsis

\input epsf
\overfullrule=0pt
\autoparens
\paper
\def\({\left (}
\def\){\right )}
\def\[{\left [}
\def\]{\right ]}
\def \lb{\left \{}
\def \rb{\right \}}
\def\ds{\displaystyle}
\def \para#1,#2,{{\ds \partial #1\over \ds \partial #2}}
\def \parb#1,#2,#3,{{\partial #1\over \partial #2}\left.\right\vert_#3}
\def \parc#1,#2,#3,{{\partial^2 #1\over\partial #2\partial #3}}
\def \pard#1,#2,{{\ds\partial^2 #1\over\ds \partial {#2}^2}}
\def\<{\left <}
\def\>{\right >}

\def\lket#1|{\langle#1|}
\def\rket#1|{|#1\rangle}
\def\kets<#1|#2>{\langle #1|#2\rangle}

\def\kp{{k^\prime}}

\def\kp{{k^\prime}}

\def\agt{\kern .2 em\raise .22em \hbox{$>$} \kern -.74em \lower.25em
\hbox{$\sim$}\kern .2 em }

\def\F{{\cal F}}

\def\F{{\cal F}}
\def\sc{\fam0\tenrm\uppercase}
\def\rr#1#2{\lparen{\sc {#2}\rparen}}
\def\rra#1#2{\lparen{\sc {#2}};}
\def\rrb#1#2{{\sc {#2}};}
\def\rrc#1#2{{\sc {#2}\rparen}}
\def\ListReferences{{\sc{Ashurst, W.~T.}}, and {\sc{Hoover, W.~G.}}(1976) Microscopic
fracture studies in the two-dimensional triangular lattice,
 {\sl  Phys.~Rev.}   {\bf B14}, 1465-1473.

{\sc{Atkinson, W.}}, and {\sc{Cabrera, N.}}(1965) Motion of a
Frenkel-Kontorova dislocation in a one-dimensional crystal, {\sl
Phys.~Rev.}  {\bf 138}, A763-A766

{\sc{Bergkvist, H.~}}(1973) The motion of a brittle crack, {\sl
J.~Mech.~Phys.~Solids}    {\bf 21}, 229-239.

{\sc{Bergkvist, H.~}}(1974) Some experiments on crack motion and
arrest in polymethylmethacrylate,, {\sl  Engng.~Fracture~Mech.}   {\bf
6}, 621-626.

{\sc{Celli, V.}}, and {\sc{Flytzanis, N.}}(1970) Motion of a screw
dislocation in a crystal, {\sl  J.~Appl.~Phys.} {\bf 41}, 4443-4447.

{\sc{ Cotterell, B.}} (1965) Velocity effects in fracture propagation,
{\sl Appl.~Mater.~Res.} {\bf 4} 227-232.

{\sc{D\"oll, W.}}, and {\sc{Weidmann, G.~W.}}(1976) Transition from
slow to fast crack propagation in PMMA, {\sl J.~Mater.~Sci.~Lett.}   {\bf 11},
2348-2350.

{\sc{Doyle, M.~}}(1983)  A mechanism of crack
branching in polymethylmethacrylate and the origin of the bands on the
surfaces of fracture, {\sl  J.~Mater.~Sci.} {\bf 18}, 687-702.

{\sc{Fineberg, J., Gross, S., Marder, M.}} and {\sc{Swinney, H.}}, {\sc{{\it \lowercase{et.~al.}}}}(1991)
Instability in dynamic fracture, {\sl
Phys.~Rev.~Lett.}  {\bf 67}, 141-144.

{\sc{Fineberg, J., Gross, S., Marder, M.}} and {\sc{Swinney, H.}}, {\sc{{\it \lowercase{et.~al.}}}} (1992)
Instability in the propagation of fast cracks, {\sl
Phys.~Rev.}   {\bf B45}, 5146-5154.

{\sc{Freund, L.~B.}}(1974)Crack
propagation in an elastic solid subjected to general loading. IV.
Obliquely incident stress pulse,  {\sl  J.~Mech.~Phys.~Solids}  {\bf
22}, 137-146. 

{\sc{Freund, L.~B.}} (1990){\it Dynamic Fracture Mechanics}
Cambridge University Press, New York.

{\sc{Fuller, K.~N.~G.}}, {\sc{Fox, P.~G.~}}, and {\sc{Field, J.~E.~}}
(1975) The temperature rise at the tip of fast-moving cracks in glassy
polymers, {\sl Proc.~R.~Soc.~Lond.~A}   {\bf 341}, 537-557.

{\sc{Gao, H.~}}(1993) Surface roughening and branching instabilities
in dynamic fracture, {\sl  J.~Mech.~Phys.~Solids}   {\bf 41}, 457-486.

{\sc{Gilman, J.~J.}}, {\sc{Knudsen, C.}}, and {\sc{Walsh, W.~P.~}}
(1958) Cleavage cracks and dislocation in LiF crystals, {\sl
J.~Appl.~Phys}   {\bf 29}, 601-607.

{\sc{Gradshteyn I.~S.}} and {\sc{Ryzhik, I.~M.}} (1980) Table of
integrals, series and products. Academic Press, New York.

{\sc{Green, A.~K.}}, and {\sc{Pratt, P.~L.}}(1974) Measurement of the
dynamic fracture toughness of polymethylmethacrylate by high-speed
photography, {\sl   Engng.~Fracture Mech.}  {\bf 6}, 71-80.

{\sc{Gross, S., Fineberg, J., McCormick, W.~D., Marder, M.,}} and
{\sc{Swinney, H.}} (1993) Acoustic
emissions from rapidly moving cracks, {\sl 
Phys.~Rev.~Lett.}  {\bf 71}, 3162-3165.

{\sc{Jackson, D.~A.}}, {\sc{Pentecost, H.~T.~A.}}, and {\sc{Powles,
J.~G.~}} (1972) Hypersonic absorption in amorphous polymers by light
scattering,  {\sl Mol.~Phys.}   {\bf 23}, 425-432.

{\sc{Kanninen, M.~F.}}, and {\sc{Popelar, C.}}(1985) {\it Advanced 
Fracture Mechanics}  Oxford University Press, New York.

{\sc{Katsamanis, F.~G.}}, and {\sc{Delides, C.~G.}}(1988) Fracture
surface energy measurements of PMMA: a new experimental approach,  {\sl
J.~Phys.}   {\bf D 21}, 79-86.

{\sc{Knauss, W.~G.}}(1966) Stresses in an infinite strip containing a
semi-infinite crack, {\sl  J.~Appl.~Mech.} {\bf 33}, 356-362.

{\sc{Kobayashi, A.}}, {\sc{Ohtani, N.}}, and {\sc{Sato, T.~}} (1974)
Phenomenological aspects of viscoelastic crack propagation, {\sl
J.~Appl.~Polymer Sci.}   {\bf 18}, 1625-1638.

{\sc{Kulakhmetova, Sh.~A.}},  {\sc{Saraikin, V.~A.}}, and
{\sc{Slepyan, L.~I.}} (1984) Plane problem of a crack in a lattice,
{\sl Mechanics of Solids}   {\bf 19}, 101-108.

{\sc{Langer, J.~S.~}}(1993) Dynamic model of onset and propagation of
fracture, {\sl Phys.~Rev.~Lett.}   {\bf 70}, 3592-3594.

{\sc{Langer, J.~S.}}, and {\sc{Nakanishi, H.~}}(1993) Models of crack
propagation. II. Two-dimensional model with dissipation on the
fracture surface, {\sl Phys.~Rev.}   {\bf E48}, 439-448.

{\sc{Liu,  X.}} (1993) Dynamics of fracture propagation,  (Dissertation,
University of Texas)

{\sc{Liu, X.}}, and {\sc{Marder, M.}}(1991) The energy of a
steady-state crack in a strip, {\sl  J.~Mech.~Phys.~Sol}
{\bf 39}, 947-961.

{\sc{Machov\'a, A.}} (1992) Molecular dynamic simulation of microcrack
initiation,  {\sl Materials Science and Engineering}
{\bf A149} 153-165.

{\sc{Manneville, P.}} (1990) {\it Dissipative Structures and Weak Turbulence}, 
Academic Press, Boston Chapter 6, section 6

{\sc{Marder, M.~}}(1991) New dynamical equation for cracks, {\sl
Phys.~Rev.~Lett.}   {\bf 66}, 2484-2487.

{\sc{Marder, M.}}, and {\sc{Liu, X.~}}(1993) Instability in lattice
fracture, {\sl  Phys.~Rev.~Lett.} {\bf 71}, 2417-2420.

{\sc{Mecholsky, J.~J.}} (1985) Fracture analysis of glass surfaces,
in {\it Strength of Inorganic Glass}, ({\sc{C.~R.Kurkjian}}, ed.),
pp.~569-590. Plenum Press, New York.

{\sc{Noble, B.}}  (1959) {\it Methods Based on the 
Wiener-Hopf Technique  for the Solution of Partial 
Differential Equations} Pergamon Press, New York.

{\sc{Perrin, G.}},  and {\sc{Rice, J.~R.}} (1994), Disordering of a
dynamic planar crack front in a model elastic medium of randomly
variable toughness, {\sl J.~Mech.~Phys.~Solids} {\bf 42} 1047-1064.

{\sc{Ravi-Chandar, K.~}}, and {\sc{Knauss, W.~G.~}}(1984) An
experimental investigation into dynamic fracture: III. On steady state
crack propagation and crack branching, {\sl
Int.~J.~Fracture}   {\bf  26 }, 141-154.

{\sc{Rice, J.~R., Ben-Zion, Y.}},  and {\sc{Kim, K}}. (1994) Three-dimensional
perturbative solution for a dynamic planar crack moving unsteadily in
a model elastic solid,  {\sl J.~Mech.~Phys.~Solids} {\bf 42}, 813-843.

{\sc{Sieradzki, K., Dienes, G.~J., Paskin, A.,}} and
{\sc{Massoumzadeh, B.}}, (1989) 
Atomistics of crack branching, {\sl Acta.~Metall.} {\bf 36}, 651

{\sc{Slepyan, L.~I.}} (1981) Dynamics of brittle fracture in lattice,
{Doklady Soviet Phys.} {\bf 26} 538-540.

{\sc{Slepyan, L.~I.}}(1992) Principle of maximum energy dissipation rate
in crack dynamics, {\sl  Doklady Soviet Phys.}   {\bf 41}, 1019-1033.

{\sc{Slepyan, L.~I.}}(1993) Principle of maximum energy dissipation rate
in crack dynamics, {\sl  J.~Mech.~Phys.~Solids}   {\bf 41}, 1019-1033.

{\sc{Takahashi, K.}}, {\sc{Matsushige, K.~}} and {\sc{Sakurada, Y.~}},
(1984) Precise evaluation of fast fracture velocities in acrylic
polymers at the slow-to-fast transition, {\sl  J.~Mater.~Sci.}  {\bf
19}, 4026-4034.

{\sc{Thomson, R.~}}(1986) Physics of fracture, {\sl  Solid State
Physics}   {\bf 39}, 1-129.

{\sc{Thomson, R.}}, {\sc{Hsieh, C.}}, and {\sc{Rana, V.}}(1971)
Lattice trapping of fracture cracks, {\sl J.~Appl.~Phys.} {\bf 42},
3154-3160. 

{\sc{Washabaugh, P.~D.}}, and {\sc{Knauss, W.~G.}}(1993) Non-steady
periodic behavior in the dynamic fracture of PMMA,  {\sl
Int.~J.~Fracture}   {\bf 59}, 189-197.

{\sc{Willis, J.~R. }}  (1990) Accelerating cracks and related
problems, , in {\it Elasticity: Mathematical Methods
and Applications} ({\sc{G.~Eason, and R.~W.~Ogden}}, eds.)
pp.~397-409. Halston Press, New York.

{\sc{Xu, Y.}}, and {\sc{Keer, L.~M.}}(1992) Non-planar deviation of an
initially straight moving crack,  {\sl Engng.~Fracture~Mech.} {\bf
41}, 577-585. 

{\sc{Yoffe, E.~H.}}(1951) The moving Griffith crack, {\sl  Phil.~Mag.}
{\bf  42}, 739-750.

{\sc{Zhou, S.~J., Carlsson, A.~E.}}, and  {\sc{Thomson, R.}}(1994) Crack
blunting effects on dislocation emission from cracks,  {\sl
Phys.~Rev.~Lett.}   {\bf 72}, 852-855.

}
\singlespace

\hyphenation{slows}
\title
Origin of Crack Tip Instabilities
\endtitle
\author
M.~Marder and Steve Gross
Department of Physics and Center for Nonlinear Dynamics
The University of Texas, Austin TX 78712
\endauthor
\abstract This paper demonstrates that rapid fracture
of ideal brittle lattices naturally involves phenomena long seen in
experiment, but which 
have been hard to understand from a continuum point of view. These
idealized models do not mimic realistic
microstructure, but can be solved exactly and understood completely.
First it is shown that constant
velocity crack solutions do not exist at all for a range of velocities
starting at zero and ranging up to about one quarter of the shear wave
speed. Next it is shown that above this speed cracks are by and large
linearly stable, but that at sufficiently high velocity they become unstable
with respect to a nonlinear micro-cracking instability.  The way this
instability works itself out is related to 
the scenario known as intermittency, and the basic time scale which
governs it is the inverse of the amount of dissipation in
the model. Finally, we compare the theoretical framework with some new
experiments in Plexiglas, and show that all qualitative features of
the theory are mirrored in our experimental results.
\endabstract
{\tenpoint Submitted to {\it Journal of the Mechanics and Physics of Solids}}
\section{Introduction}
The classic theory of fracture \rra{Freund90}
{Freund 1990}
\rrc{Kan}
{Kanninen and Popelar 1985}
 treats cracks as
mathematical branch 
cuts which begin to move when an infinitesimal extension of the crack
releases more energy than is needed to create fracture surface. This
idea is enormously successful in practice, but conceptually incomplete.
We will show that one cannot  understand how a crack moves
without  taking into account
the fact that it moves in a fundamentally discrete and not continuous medium.
In a lattice 
there are some  velocities for which crack solutions do not exist
at all, others where cracks are linearly unstable and accelerate to
higher velocities, and others for which crack tips are unstable and
break apart altogether. While these conclusions are all compatible
with continuum mechanics, they were not predicted by it, and are
somewhat surprising.

The experimental observations which motivate this study are 
\itemize
\itm that cracks in amorphous brittle materials such as glass or
Plexiglas  pass almost instantaneously from quasi-static motion to
motion at about 10\% of the Rayleigh wave
speed \rra{Doll_Transition}
{D\"oll and Weidmann 1976}
\rrc{Takahashi}
{Takahashi, Matsushige, and Sakurada 1984}
. 
\itm that they seldom travel faster than 60\% of the Rayleigh wave
speed \rra{Kobayashi}
{Kobayashi, Ohtani, and Sato, 1974}
\rrc{Ravi}
{Ravi-Chandar and Knauss, 1984}, although
according to continuum theory \rr{Freund90}
{Freund, 1990}
 this wave speed should be
the limiting velocity.
\itm that at about 40\% of the Rayleigh wave speed, the acceleration
of cracks slows sharply, their velocity begins to
oscillate
\rra{Fineberg_91}
{Fineberg {\lowercase {\it et.~al.}}, 1991, 1992}
\rrc{Knauss93}
{Washabaugh and Knauss, 1993}
, they emit high-frequency acoustic
waves \rr{Gross_93}
{Gross {\lowercase {\it et.~al.}}, 1993}
, the energy needed to form new crack
surface increases dramatically
\rr{Pratt_and_Green}
{Green and Pratt, 1974}
, and the fracture
surface shows periodic 
structure \rr{Mecholsky}
{Mecholsky, 1985}
 correlated with velocity
oscillations \rr{Fineberg_92}
{Fineberg, {\lowercase {\it et.~al.}}, 1992}
. The 
basic time scale of the  velocity oscillations is much larger than the time
scale on which the crack snaps atomic bonds, and has been unexplained.
\enditemize
The continuum theory of fracture contains many hints concerning these
phenomena, but its predictions have always seemed ambiguous. In the
very first detailed calculation concerning dynamic fracture,
Yoffe
\rr{Yoffe}
{1951} showed that at around 60\% of the Rayleigh wave speed,
a crack should become unstable, since the maximum tensile stress would
no longer be directly ahead of the crack, but would instead be off at
an angle. The dynamical implications of this calculation were however
unclear. Would the crack branch,  would it begin oscillatory motion,
or do something else?
To make matters more puzzling, depending on precisely which component
of the stress one monitors, the instability can set in at different
velocities, and there is no criterion deciding between the various
possibilities \rr{Freund90}
{Freund 1990}. For example, Freund \rr{Freund_74}
{1974} has found
that the velocity at which a 
moving crack consumes enough energy so that  two nearly static
cracks could travel instead of one fast one is 45\% of the Rayleigh
wave speed $c_R$, while  
some recent perturbative calculations \rra{Xu_92}
{Xu and Keer, 1992}
\rrc{Gao_93}
{Gao, 1993}
find a dynamical
instability at around 65\% of $c_R$. In addition there are hints of
instabilities in calculations coupling thermal and mechanical
motion
\rra{Langer_93a}
{Langer, 1993a}
\rrc{Langer_93b}
{Langer and Nakanishi, 1993}
, and in other simple energy balance
arguments
\rra{Slepyan_92}
{Slepyan, 1992}
\rrb{Gao_93}
{Gao, 1993}
\rrc{Slepyan_93}
{Slepyan, 1993}
. 

All of these calculations are
constrained by the fact that continuum elasticity is not well suited to
describe the microscopic processes by which elastic energy is
converted to broken bonds. We believe that understanding how
this happens, even in the simplest case,  is the key to resolving  the
qualitative puzzles mentioned above
\rra{Marder.PRL.93}
{Marder and Liu, 1993}
\rrc{Liu_93}
{Liu, 1993}. 

The idealized
models of brittle lattice fracture we will solve in this paper are not
intended to describe material microstructures realistically, but,
since they can be solved analytically,  provide solid ground on
which to show how the dynamics of fracture alter when discreteness is
taken into account.  In these models, steady state crack
motion is impossible for velocities less than around 30\% of the
relevant wave speed, and at around 50\% of this wave speed, steady state crack
motion becomes unstable. We will describe the precise dynamical way
the instability occurs, show some of its consequences after onset, and
compare the results with experiment. 

The type of instability found in the lattice theory is similar to that
known by the term  intermittency
\rr{Manneville}
{Manneville, 1990}, and it occurs in a simple way. The
crack tip starts from some initial condition, and 
tries to settle into a steady state configuration in which it snaps
one by one the atomic bonds perpendicular to its direction of motion.
Just as it is about to reach this state, a seemingly irrelevant bond
parallel to the direction of motion snaps and throws the crack off its
course. The crack tip is thrown far away from the steady state
configuration, and gathers itself up to try the approach again,
repeating the process periodically. One of the surprises in the
analysis is that the time scale which governs the approach of the
crack tip to the steady state is given by the time for transient
perturbations to die out, and is formally just the inverse of the
amount of dissipation in the model. Physical systems with small
amounts of dissipation should be expected to show oscillations on
relatively long time scales. This idea provides a tentative solution
to the most perplexing experimental observation.

From a formal point of view the calculations described here build upon
the work of Slepyan \rr{Slepyan}
{1981} and co-workers
\rr{Kulamekhtova}
{Kulamekhtova, Saraikhin, and Slepyan,  1984}
, lattice solutions for dislocations, \rra{Atkinson}
{Atkinson and Cabrera, 1965}
\rrb{Celli}
{Celli and Flytzanis, 1970}
\rrc{Thomson}
{Thomson, Hsieh, and Rana, 1971}
and upon results of prior
numerical simulations \rra{Ashurst}
{Ashurst and Hoover, 1976}
\rrb{Sieradzki}
{Sieradzki { \lowercase {it et.~al.}}, 1988}
\rrc{Machova}
{Machov\'a, 1992}
. An interesting
new result is that many of the steady state configurations
which have been derived 
previously are unphysical and do not exist. In addition, the stability
of the remaining steady states is examined carefully. The general rule
is that any state whose velocity increases when one pulls on it harder
is linearly stable; however, there are nonlinear instabilities to
watch for as well, and the points where these occur
are determined through a combination of analytical and
numerical techniques. The smallest negative eigenvalue governing
approach of transient configurations to the steady state is calculated
and found to be simply the inverse of the coefficient of dissipation
in the model.

There are some clear differences between the behavior of the lattice models
and experimental systems. The velocities at which various processes
occur in the models are all distinctly higher than their supposed
counterparts in experiment. Whether these differences can be
attributed to the fact that the model is schematic, or whether
additional physical processes altogether actually operate in the
experiment, will have to be determined in the future. Some of the most
important additional processes to consider have to do with the fact
that the experiments are conducted in polymeric solids, not simple
lattices, and are fully three-dimensional
\rra{Rice_94a}
{Rice, Ben-Zion, and Kim, 1994}
\rrc{Perrin_94}
{Perrin and Rice, 1994}
 while the
theory considers only two dimensions. The solutions considered
in this paper do not allow dislocations \rr{Zhou}
{Zhou, {\lowercase{\it et.~al.}}, 1994}
. This fact may curiously make
comparison with polymers more appropriate than with crystals, despite
the fact that the calculations are carried out in lattices.

Because analytical calculations of crack motion in lattices are
elaborate, the paper will proceed in steps. First, in Section
\use{sec.continuum} we will review the theory of crack motion in a
continuum strip. Next, in Section \use{sec.1D} we  will present a
one-dimensional model which displays almost 
all the features of the more elaborate cases. In Section \use{sec.modeIII}
we will solve a two-dimensional model in which
mass points move with only one degree of freedom (Mode III).
In Section \use{sec.modeI} the mass points will be allowed
to move with two degrees of freedom (Mode I), and the the calculations
compared with some new experimental results  in Section \use{sec.exp}.

\section{Crack motion in a continuum strip\label{sec.continuum}}

\figure{strip}
\infiglist{Semi-infinite crack moving in an infinite strip}
\Caption
All the calculations in this paper will concern a semi-infinite crack
moving through the center of an infinite strip.
\endCaption
\epsfxsize=6 truein
\centerline{\epsffile{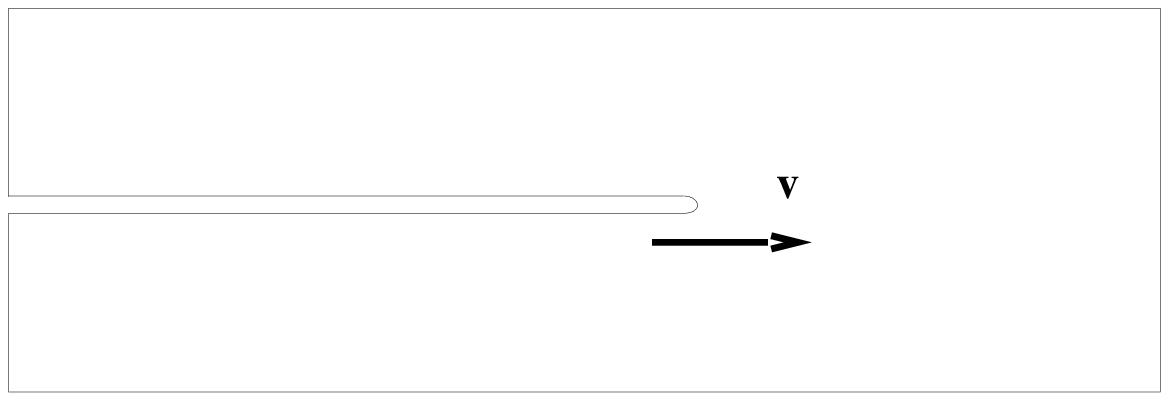}}
\endfigure
The successes of dynamic fracture mechanics \rr{Freund90}
{Freund, 1990}
 provide the
backdrop to this paper. The most complete understanding of
crack dynamics in a continuum has been obtained for a semi-infinite
crack in an infinite plate, with the crack driven by loading on its
faces \rr{Willis}
{Willis, 1990}
. However, we will study a different geometry here, that of a
semi-infinite crack in an infinite strip \rra{Knauss_66}
{Knauss, 1966}
\rrb{Bergkvist_73}
{Bergkvist 1973}
\rrc{Bergkvist_74}
{Bergkvist 1974}
. There are two
reasons for the choice. First, the choice of a strip makes it simpler
to compare 
with numerical simulations over long periods of time, since the
effects of top and bottom boundaries are included in the calculation.
Second, the strip geometry is the only one in which time-independent
loading naturally leads a crack to move in steady state at a constant
velocity. In an infinite plate, generic time-independent loading
causes a crack to accelerate indefinitely.

What is known about crack motion in a strip (\Fig{strip})? 
In certain respects, the geometry is extremely convenient, since
when a crack
moves some distance $l$ along the middle of a strip, all features of
the problem are unchanged except that stressed material in advance of
the crack has been converted to unstressed material behind it. For
this reason, if elastic energy $W$ is stored per unit length ahead of
the crack, then the energy flowing to a steadily moving crack tip to create new
surface is also $W$ per unit length.  The expected
asymptotic solution in a strip is that the crack will move at the velocity
$v$ such that
$$W=\Gamma(v),\EQN L93.0.0$$
where $\Gamma(v)$ is the fracture energy per unit length crack advance.
There is an apparent problem with this picture.
What happens if $\Gamma(v)$ is constant, but $W$
is slightly more than $\Gamma$? 
In this event, the crack absorbs the excess 
energy by a continual slow acceleration, moving adiabatically up
through steady states and asymptotically approaching a terminal
velocity given by a the Rayleigh wave speed \rra{MarderPRL}
{Marder, 1991}
\rrc{Liu1}
{Liu and Marder, 1991}
. An
approximate 
equation of motion which describes such  cases is \rr{Liu1}
{Liu and Marder, 1991}
$$\dot v ={c_l^2\over w} [1-\Gamma(v)/W](1-v^2/c_R^2)^2\EQN
L93.0.1$$
with $w$ the half-width of the strip, $c_R$ the Rayleigh wave speed, and
$c_l$ the longitudinal
wave speed.
The right hand side of \Eq{L93.0.1} is a simple approximation of
elaborate analytical expressions; it is
only accurate to  around 10\%, and that only when the dimensionless group
$\dot vw/c_l^2$ is small, but is very useful for qualitative analysis.

The particular form of $\Gamma(v)$ is not specified by continuum
elasticity, and by varying $\Gamma(v)$ one has the potential to find a
wide range of fracture dynamics. Experiments show a very rapid jump in
velocity at the onset of fracture, a fact we will explain later as a
consequence of the lattice theory, but one can also explain it
in the context of the continuum by studying  a case in which fracture energy
initially decreases with velocity, and then increases again. Taking
$$\Gamma(v)=\Gamma(0)-W_0({v\over 2 c_R}-({v\over c_R})^2)\EQN L93.0.2$$
leads to the fracture dynamics  shown in \Fig{vel}. States at $v=0$
for $W$ slightly less than $\Gamma(0)$ are actually metastable, since
crack motion would be possible if it could be initiated.

\figure{vel}
\infiglist{Velocity of crack in the continuum}
\Caption
Solution of \Eq{L93.0.1} with $\Gamma$ given by \Eq{L93.0.2} and shown
in inset, for $W_0=1$, $\rho=1$, $c_R=.9$, $w=1$, and
$v(t=0)=1\times 10^{-6}$. A long period in which the crack struggles to move
is followed by a period of rapid acceleration, and a final
approach to steady motion.
\endCaption
\epsfxsize=5 truein
\centerline{\epsffile{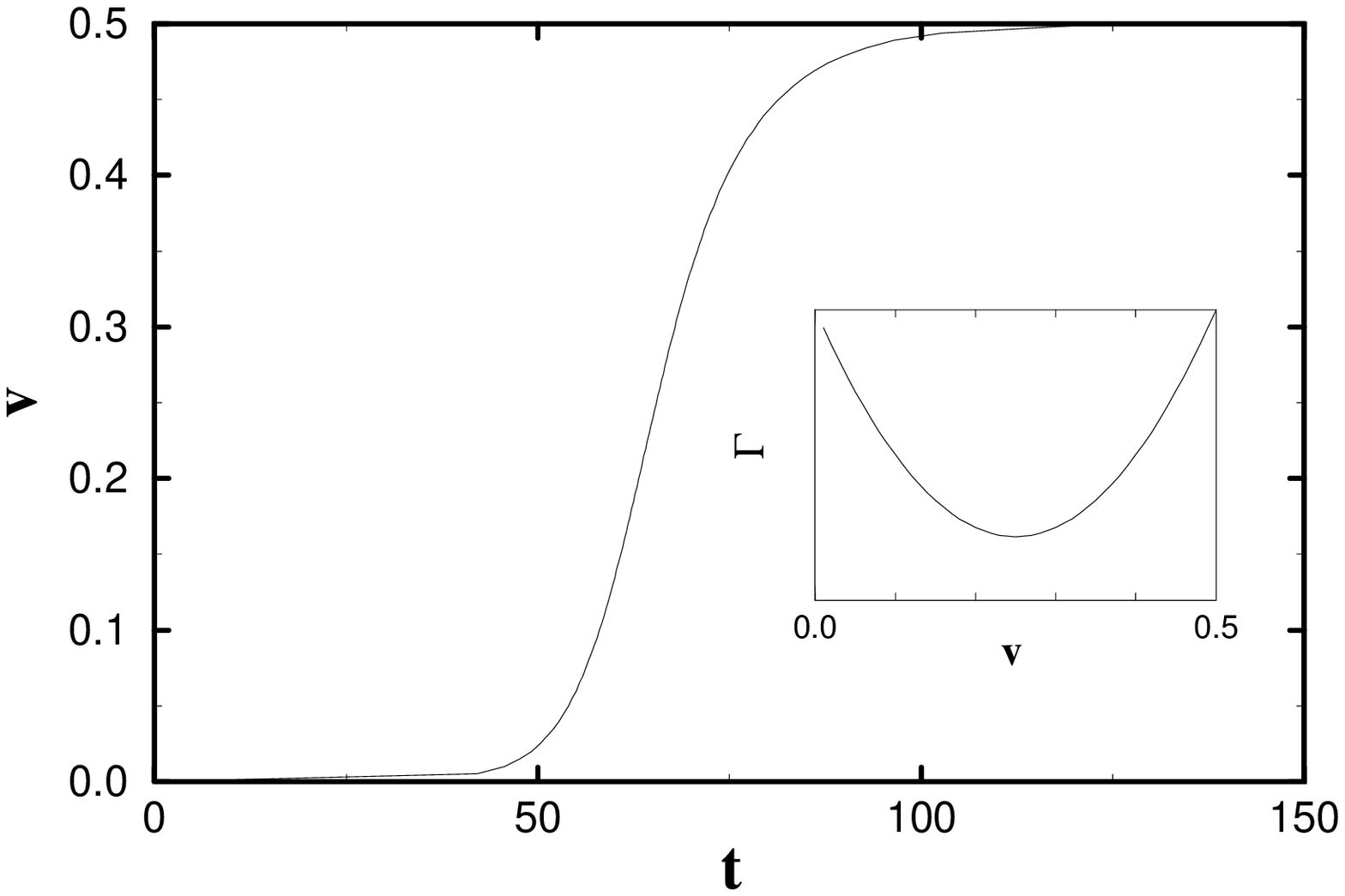}}
\endfigure
\figure{vel2}
\infiglist{Steady States}
\Caption
Solution of \Eq{CONT5} with $\Gamma(0)/W_0=1$ and $c_R=1$.  Only the
solutions indicated by the thick line are stable.
\endCaption
\epsfxsize=2.5 truein
\centerline{\epsffile{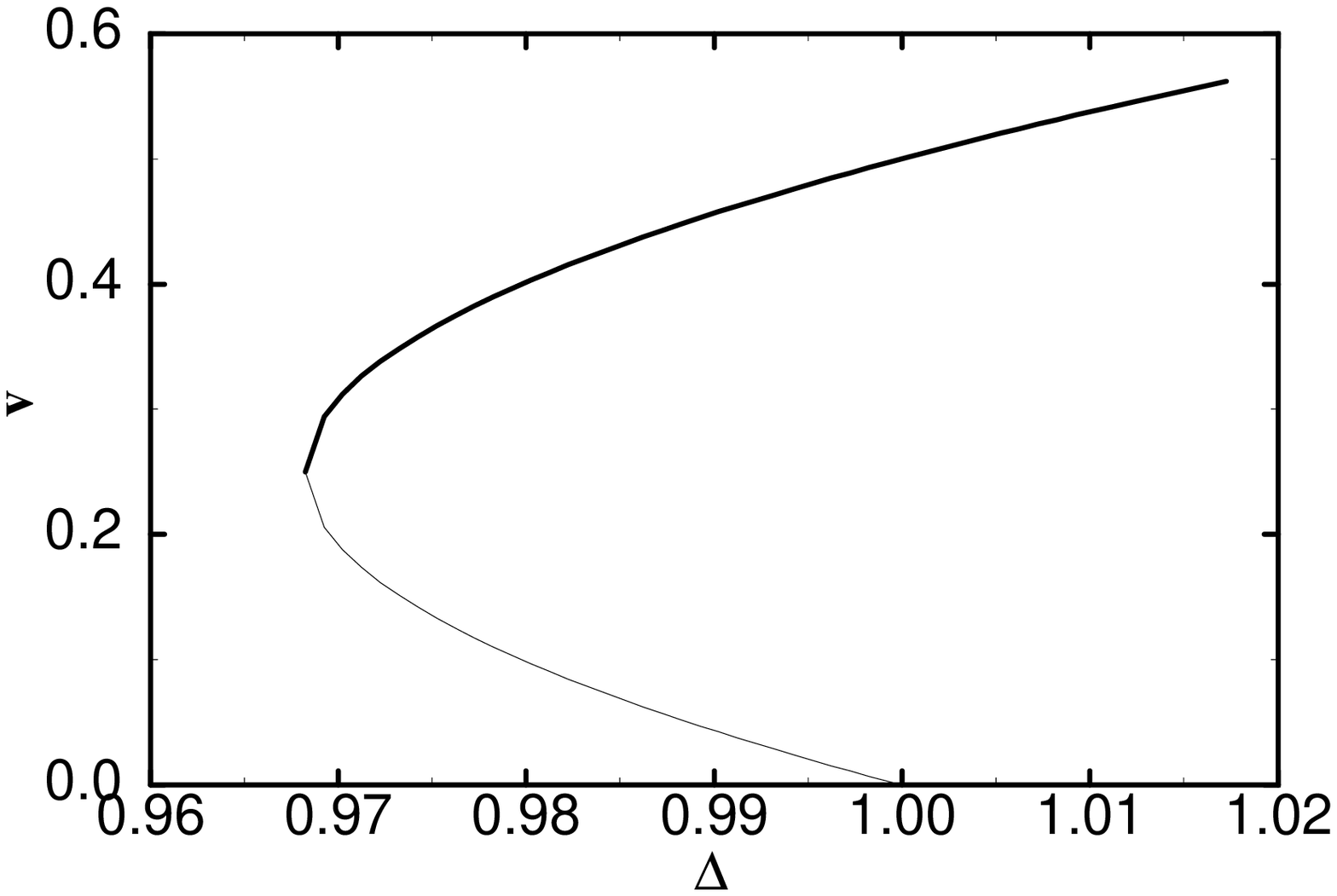}}
\endfigure

Another way to look at the case in which fracture energy $\Gamma$
initially decreases as a function of $v$ is to ask what steady
velocities are possible as a function of the energy stored in the
strip far ahead of the crack. In order to make a connection with
results to be obtained in lattices, define
$$\Delta=\sqrt{W/\Gamma(0)},\EQN CONT4$$
so that $\Delta$ is proportional to the vertical displacement of the
sides of the strip ahead of the crack. Then solving \Eq{L93.0.0} and
\Eq{L93.0.2} for $v/c_R$ gives
$$\sqrt{{\Gamma(0)\over W_0}(\Delta^2-1)+{1\over 16}}+{1\over
  4}={v\over c_R},\EQN CONT5$$
which is pictured in \Fig{vel2}. The point of this calculation is to
show that if fracture energy is initially a decreasing function of
velocity, there must be a forbidden band of steady state velocities;
cracks which start on the lower branch shown in \Fig{vel2} quickly
accelerate to the upper one, meaning that such fractures are
activated. 

Thus within continuum theory one may have a
range of velocities in which cracks are unstable and accelerate
rapidly towards some minimum stable velocity. However, lattice
theories predict something even more severe: that there is a range of
velocities in which steadily moving solutions do not exist at all.

\section{One-dimensional model\label{sec.1D}}
\figure{balls}
\infiglist{Drawing of One-dimensional Model}
\Caption
This one-dimensional model mimics the motion of a crack in a strip,
incorporating effects of discreteness. One can view it as a model for
the atoms lying just along the surface of a crack. The mass points are
only allowed to move vertically, and are tied to their neighbors with
springs which break when they exceed a certain extension. The lower portion
of the figure shows an actual solution of the model, using \Eq{L93.6},
at $v=0.5$, b=0.01.
\endCaption
\epsfxsize=5 truein
\centerline{\epsffile{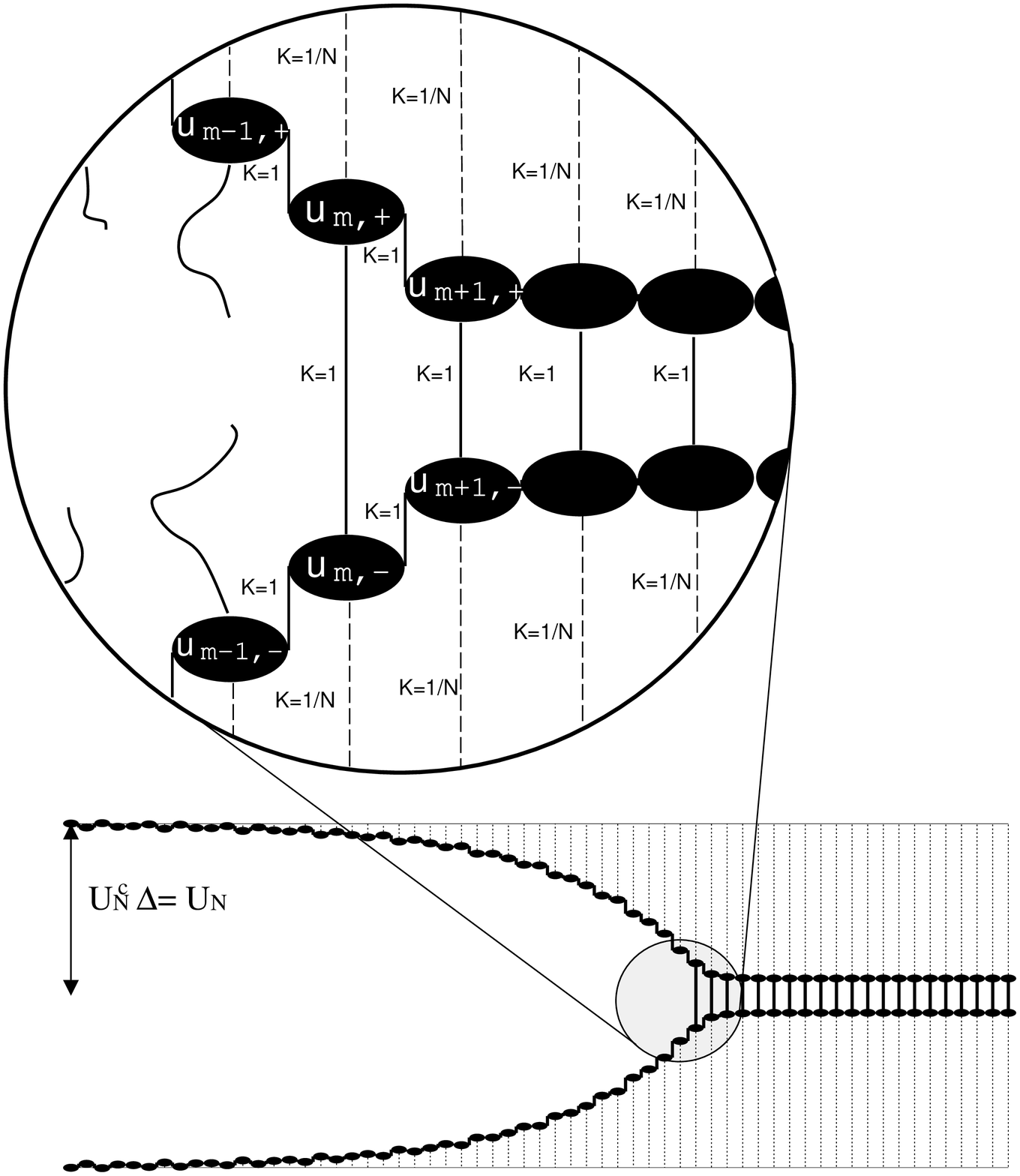}}
\endfigure
\subsection{Definition and Energetics}
In order to mimic the motion of a crack in a strip, including the
effects of underlying discreteness, but otherwise making the
calculation as simple as possible, consider the model shown in
\Fig{balls}. One 
can view it as a model for the atoms lying just along a crack
surface. They are tied to  nearest neighbors by elastic springs, with
spring constant $K=1$,
and tied to a line of atoms on the other side of the crack line by
similar springs, which however snap when extended past some breaking
point. The lines of atoms are being pulled apart by weak springs of
spring constant $K=1/N$. These weak springs are meant schematically to
represent $N$ vertical rows of atoms pulling in series, and in later
sections will be treated more realistically. The equation which
describes the upper row of mass points in this  model is
$$\ddot u_{m,+}=
\cases{u_{m+1,+}-2u_{m,+}+u_{m-1,+}&\sl  Elastic coupling to neighbors \cr 
+{1\over N}(U_N-u_{m,+}) &\sl  Driven by displacing edges of  strip\cr 
+(u_{m,-} -u_{m,+})\theta(2u_f-|u_{m,-}-u_{m,+}|)& \sl Bonds which snap\cr
-b\dot u_{m,+}& \sl Dissipation.
}\EQN OD1$$

There are a few terms that need discussion. 
First, $\theta$ is a step function, and the term containing it
describes bonds which snap when their total extension reaches a
distance $2u_f$, where $u_f$ is a fracture distance.
Second, a small
amount of dissipation has been added, the term proportional to $b$.
Originally this term was included for formal reasons in order to make
Fourier transforms well defined, but it will eventually turn out to
have physical importance. The amount of dissipation will usually be
taken to be vanishingly small. 
Third, the height that mass points
reach after the crack has passed is $U_N$, and this term
describes the force driving crack motion. 

In the discussion that follows, some of the equations will have boxes
to the left,
like \Eq{OD1.1}. These are equations that are true regardless of the
particular lattice that is being considered, and will be taken over
without change in later sections.

As $N$ varies, the natural scale on which a displacement $U_N$ is able
to drive crack motion varies, so one should find the natural
dimensionless constant which governs crack motion. The question to ask
is, { when is enough energy stored in the strip, per lattice spacing
far to the right of the crack, to break one bond along the crack
line?} An important physical quantity to define in answering this
question is obtained by going far to the right of the crack, and
taking the ratio of the displacement of the atom just above the
crack line, $U_{\rm right}$ to the total displacement at the top of
the strip $U_N$. Denote 
this ratio by
\def\QED{\vskip .2truein\line{\vrule height 1.8ex width 2ex depth
    +.2ex\hfill}\vskip -.47truein}
\QED
$$
 Q_0\equiv{\strut\ds U_{\rm right}\over \ds \strut U_N}.
\EQN OD1.1$$
Suppose that mass points far to the left and far to the right of the
crack are stationary, and that dissipation is negligible. 
Far to the right of the crack, one 
has that 
$U_{\rm right}=u_{m+}=-u_{m-}$, and  so balancing forces on masses
with $m\gg 0$ gives
$$2 U_{\rm right}={1\over N}(U_N-U_{\rm right})\EQN OD2$$
$$\Rightarrow Q_0={1\over 2N+1}.\EQN OD3$$
Remembering a factor of two for the upper and lower halves of the
strip, the energy per bond for $m\gg 0$ is therefore
$$2{1\over 2} {1\over N} (U_N-U_{\rm right})^2+{1\over
  2}(2U_{\rm right })^2\EQN OD5$$
$$
\Rightarrow E_{\rm right}=2Q_0(U_N)^2.
\EQN OD6$$
Far to the left of the crack, one simply has that
\QED\vskip .1truein
$$\strut U_{\rm left}=U_N.
\EQN OD4$$
The energy per bond for $m\ll 0$ is
just that needed to stretch the spring between
$u_{m+}$ and $u_{m-}$ from rest to breaking, and is just 
$$ E_{\rm break}=2 u_f^2 \EQN OD4.1$$
Assuming there is no other sink of energy,  one must have that
$$2Q_0{(U_N)^2}=2u_f^2.\EQN OD7$$
Therefore, the minimum value of $U_N$ at which enough energy is stored
to the right of the crack so as to be able to snap the bonds along the
crack line is
\QED
$$\strut U_N^c={\ds \strut  u_f\over \ds\strut  \sqrt{Q_0}}.\EQN OD8$$
For this reason, it is convenient to define a dimensionless measure of
how far one has pulled the edges of the strip, 
\QED
$$\Delta\equiv {\ds \strut U_N\sqrt{Q_0}\over\strut \ds u_f}.\EQN OD8.0$$
By definition, energy balance requires that steady crack motion is
only possible for $\Delta\geq 1$, and a system in which a crack moved
when $\Delta=1$ would have to be perfectly efficient in turning
potential energy into crack motion.

\subsection{Wiener-Hopf solution for steady states}
While the preceding calculations motivate the definition \Eq{OD8},
they are based upon false assumptions. When a crack moves in steady
state, Slepyan \rr{Slepyan}
{1981} first showed that the mass points far from
its tip are necessarily in  
motion. As a result the energy accounting carried out above is wrong,
and  crack motion is only  possible for $\Delta>1$. The goal now is to
examine steady states in detail.

A steady state in a lattice is more complicated than one in a
continuum;  it is a configuration which repeats itself after a 
time interval $1/v$, but moved over by one lattice spacing.
In steady state, one has the symmetries
$$u_{m+}=-u_{m-},\EQN OD8.1$$
$$u_{m+}(t)=u_{0+}(t-m/v),\EQN OD9$$
which means that all spatial behavior is contained in the
time history of any single mass point. Take this mass point to be
$u_{0+}$, and denote it simply by $u(t)$. Using \Eq{OD9} and
\Eq{OD8.1} in \Eq{OD1} gives
$$\ddot u=u(t-1/v)-2u+u(t+1/v)+{1\over N}(U_N 
-u) -2u\theta(2u_f-2|u|)-b\dot u.\EQN OD10$$ 

\Eq{OD10} can be solved analytically using Wiener-Hopf
methods \rr{Wiener-Hopf}
{Noble, 1959}
. Here 
is the solution. 

There must be some first time at which $u(t)$ rises to $1$  and the
$\theta$ function vanishes. Let us take this time to be $t=0$.
Assuming that $u$ rises above the height of $1$ once and remains above
it for good, one can write
$$\ddot u=u(t-1/v)-2u+u(t+1/v)+{1\over N}(U_N e^{-\alpha |t|}
-u) -2u\theta(-t)-b\dot u.\EQN OD11$$ 
\Eq{OD11} does not quite follow from \Eq{OD10}, since the factor
$\exp(-\alpha |t|)$ has appeared. It is introduced just to
make Fourier integrals converge, and $\alpha$ will tend to zero at the
end of the calculation.

Everything in \Eq{OD11} can be Fourier transformed in a
straightforward way except for the term $\theta(t) u(t)$. Simply
define
$$U^-(\omega)=\int dt e^{i\omega t} \theta(-t) u(t)\EQN OD12;a$$
and
$$U^+(\omega)=\int dt e^{i\omega t} \theta(t)u(t)\EQN OD12;b$$
so that $U(\omega)$, the Fourier transform of $u(t)$, is
$$U(\omega)=U^+(\omega)+U^-(\omega)\EQN OD13$$
The crucial observation is that $U^+$ is free of poles in the upper
half complex $\omega$ plane, while $U^-$ is free of poles in the lower
half plane, since the integrals \Eq{OD12} are obviously convergent in
these cases.

Using these definitions, one can now transform \Eq{OD11} to read
$$-\omega^2 U=2(\cos\omega/v-1)U+{U_N\over N}\lb{1\over
  \alpha+i\omega} +{1\over\alpha-i\omega}\rb -{U\over N} -
2U^-+i\omega b U.\EQN OD14$$
$$\Rightarrow U(\omega) F(\omega)-2 U^-(\omega)=-{U_N\over N}\lb{1\over
  \alpha+i\omega} +{1\over\alpha-i\omega}\rb\EQN OD15$$
with
$$F(\omega)=\omega^2+2(\cos\omega/v-1)-{1\over N}+i\omega b.\EQN
OD15.1$$
Solving for $U^-$  with the aid of \Eq{OD13}
gives 
$$U^+(\omega) {F(\omega)\over F(\omega)-2} + U^-(\omega)=-{U_N\over
  N[F(\omega)-2] }\lb{1\over
  \alpha+i\omega} +{1\over\alpha-i\omega}\rb.\EQN OD15.0.1$$
Define
$$Q(\omega)={F(\omega)\over F(\omega)-2}\EQN OD17$$
and use the fact that $\alpha$ is vanishingly small, so one only needs
the value of $F(0)=-1/N$ on the right hand side of \Eq{OD15.0.1} to
write
\QED
$$Q(\omega)U^+(\omega)+U^-(\omega)=Q_0U_N\lb{\ds \strut 1\over \ds \strut
  \alpha+i\omega} +{\ds \strut 1\over\ds\strut\alpha-i\omega}\rb.\EQN OD16$$
Here $Q_0$ is given by \Eq{OD3}, and one can easily check that
\QED\vskip .1truein
$$Q_0=Q(0).\strut\EQN OD16.0$$
From a formal point of view, \Eq{OD16} is important because all of the
lattice models in subsequent sections can be put in exactly this form,
with the function $Q$ becoming progressively more complicated as the
model becomes more realistic, but everything else remaining precisely
the same.

The Wiener-Hopf technique directs one to write
\QED
$$
Q(\omega)={\ds\strut Q^-(\omega)\over \ds\strut Q^+(\omega)},\EQN OD18$$
where $Q^-$ is free of poles and zeroes in the lower complex $\omega$
plane and $Q^+$ is free of poles and zeroes in the upper complex
plane. One can carry out this decomposition with the explicit formula
\QED
$$
\ds Q^\pm(\omega)=\exp[\mp\int dt\, e^{i\omega t} \theta(\pm t) \int
{\ds\strut d\omega^\prime\over \ds\strut 2\pi} e^{-i\omega^\prime t}
\ln Q(\omega^\prime)]^{\vphantom{0}}_{\vphantom{0}} ,
  \EQN L93a.2;a$$
\QED
$$\ds =\exp[\lim_{\epsilon\rightarrow 0}  \int
{\ds\strut d\omega^\prime\over \ds\strut 2\pi} {\ds \ln
Q(\omega^\prime)\over\ds  i\omega \mp\epsilon
  -i\omega^\prime} ]^{\vphantom{0}}_{\vphantom{0}}.  \EQN L93a.2;b$$

Now separate \Eq{OD16} into  two pieces, one of
which has poles only in the lower half plane, and one of which has poles
only in the upper half plane: 
\QED
$$ 
{\ds U^+(\omega)\over \ds  Q^+(\omega)} -
{\ds Q_0U_N\over  \ds Q^-(0)} {\ds 1\over\ds  (-i\omega+\alpha)}
={\ds Q_0U_N\over \ds  Q^-(0)} {\ds 1\over \ds(i\omega+\alpha)}-
 {\ds U^-(\omega)\over \ds Q^-(\omega)} .
\EQN L93.5$$
Because the right and left hand sides of this equation have poles in
opposite sections of the complex plane, they must separately equal a
constant, {\cal C}. The constant must vanish, or $U^-$ and $U^+$ will
behave as a delta function near $t=0$. So
\QED
$$
U^-(\omega)={U_N }{\ds\strut Q_0Q^-(\omega)\over \ds\strut Q^-(0)
(\alpha+i\omega)},\EQN L93.6;a$$
and
\QED
$$U^+(\omega)={U_N }{\ds Q_0\strut Q^+(\omega)\over \ds\strut Q^-(0)
(\alpha-i\omega)}. \EQN L93.6;b$$
One now has an explicit solution for $U(\omega)$. Numerical evaluation
of \Eq{L93a.2;a}, and $U(t)$ from \Eq{L93.6} is fairly
straightforward, using fast Fourier 
transforms. However, in carrying out the numerical transforms, it is
important to analyze the behavior of the functions for large values of
$\omega$. In cases where functions to be transformed decay as
$1/i\omega$, this behavior is best subtracted off before the numerical
transform is performed, with the appropriate step function added back
analytically afterwards. Conversely, in cases where functions to be
transformed have a step function discontinuity, it is best to subtract
off the appropriate multiple of $e^{-t}\theta(t)$ before the
transform, adding on the appropriate multiple of $1/(1-i\omega)$
afterwards. A solution of \Eq{L93.6} constructed in this manner
appears in \Fig{balls}.

\subsection{Relation between $\Delta$ and $v$}
There is an important point which has been forgotten. This solution is
only correct if in fact 
$$u(t)=u_f\ {\rm at} \ t=0\EQN OD16.5$$
because this is supposed to be the moment at which the bond between
$u_{0+}$ and $u_{0-}$ breaks. The only parameter in the problem one is
free to vary is $U_N$, so the condition \Eq{OD16.5} chooses a value
of $U_N$, or its dimensionless counterpart,  $\Delta$. Once one
assumes that the crack moves in steady state at 
velocity $v$, there is a unique $\Delta$ which makes it possible.

To obtain \Eq{OD16.5}, one needs to require that
$$\lim_{t\rightarrow 0^-}\int {d\omega\over 2\pi} e^{-i\omega t}
U^-(\omega) =u_f.\EQN L93a.3$$
This integral can be evaluated by inspection. One knows that for
positive $t>0$, 
\QED\vskip .1truein
$$\int d\omega \exp[-i\omega t] U^-(\omega)=0\EQN OD16.5.1$$ , and
that any function whose 
behavior for large $\omega$ is $1/i\omega$ has a step function
discontinuity at the origin. Therefore, \Eq{L93a.3} and \Eq{L93.6;a} become
$$u_f={U_N Q_0}{Q^-(\infty)\over Q^-(0)}.\EQN L93a.4.0$$
Since from \Eq{OD17} follows that  $Q(\infty)=1$, one sees from
\Eq{L93a.2;b} that 
\QED\vskip .1truein
$$ Q^-(\infty)=Q^+(\infty)=1 . \EQN L93a.4.1$$ 
As a result, one has from
\Eq{L93a.4.0} and the definition of $\Delta$ given in \Eq{OD8.0} that
$$\Delta={Q^-(0)\over \sqrt {Q_0}}.\EQN L93a.4$$

To make this result more explicit, use \Eq{L93a.2;b} and the fact that
$Q(-\omega)= \bar Q(\omega)$ to write
$$Q^-(0)=\exp[\int {d\omega^\prime \over 2\pi} {1
\over 2}[{ \ln Q(\omega^\prime)\over
\epsilon-i\omega^\prime}
+{ \ln Q(-\omega^\prime)\over
\epsilon+i\omega^\prime} ]]\EQN TL23$$
$$=\exp[\int {d\omega^\prime \over 2\pi} [{1\over
-2i\omega^\prime} \ln \lb{ Q(\omega^\prime)\over \bar 
Q(\omega^\prime)}\rb +{\epsilon\over
\epsilon^2+{\omega^\prime}^2} \ln Q(0)]]\EQN TL25$$
\QED
$$\ds
\Rightarrow
Q^-(0)=\sqrt{Q_0} \exp[-\int {d\omega^\prime \over 2\pi}
{1\over 2i\omega^\prime} \ln \lb{
\ds\strut Q(\omega^\prime)\over\ds\strut  \bar  
Q(\omega^\prime)}\rb ]^{\vphantom{0}}_{\vphantom{0}}.\EQN TL25.0$$
Placing \Eq{TL25.0} into \Eq{L93a.4} gives
$$\Delta=\exp[-\int {d\omega^\prime \over 2\pi}
{1\over 2i\omega^\prime} \ln \lb{ Q(\omega^\prime)\over \bar  
Q(\omega^\prime)}\rb ].\EQN TL26$$

This expression is not completely general because the fracture
condition that $U(t=0)$ must equal $u_f$ is not completely general. In
the lattice considered in Section \use{sec.modeI}, one has  instead
that $U(t=0)$ must equal $2 u_f/\sqrt 3$. Apart from this constant of
proportionality everything goes through as above, and one has the
general result that
\QED
$$\ds
\Delta=C\exp[-\int {\ds \strut d\omega^\prime \over \ds\strut 2\pi}
{\ds 1\over \ds 2i\omega^\prime}\lb \ln \ds Q(\omega^\prime)- \ds \overline { 
\ln Q(\omega^\prime)}\rb ]^{\vphantom{0}}_{\vphantom{0}}, \EQN TL26.0$$
where $C$ is a constant of order unity that is determined by the
geometry of the lattice, equaling $1$ for the model of this section
and the model of Section \use{sec.modeIII}, and $2/\sqrt3$ for the
model of Section \use{sec.modeI}. When written in this form,
\Eq{TL26.0} is suitable for numerical evaluation, since there is no
uncertainty relating to the phase of the logarithm. 

When $b$ becomes sufficiently small, $Q$ is real for real $\omega$ except in
the small neighborhood of isolated roots and poles that sit near the
real $\omega$ axis. Let $r_i^+$ be the roots of $Q$ with negative
imaginary part (since they belong with $Q^+$), $r_i^-$ the roots of
$Q$ with positive imaginary part, 
and similarly $p_i^\pm$ the poles of $Q$. Then one can rewrite
\Eq{TL26.0} as
\QED\nobreak
$$
\Delta=C\sqrt{\ds\strut \prod{\ds\strut r_i^-p_i^+\over \ds\strut
r_i^+p_i^-}}^{\vphantom{0}}_{\vphantom{0}},\ {\rm for}\ b=0.\EQN OD20$$ 

Together with \Eq{L93.6}, \Eq{TL26.0} and \Eq{OD20}  constitute the
formal solution 
of the model. Since $Q$ is a function of the steady state velocity
$v$, \Eq{TL26} relates the external driving force on the system,
$\Delta$, to the velocity of the crack $v$.

\figure{oned.energy}
\infiglist{$v$ versus $\Delta$ for the one-dimensional model}
\Caption The velocity of a crack $v$ is plotted as a function of the
driving force $\Delta$ for the one-dimensional model. The calculation
is carried out using \Eq{OD20} for $N=100$ and $v=0.5$. The thick
upper line indicates physically realizable solutions, and the line
along $v=0$ indicates the range of lattice-trapped solutions.
\endCaption
\epsfysize=3.5 truein
\centerline{\epsffile{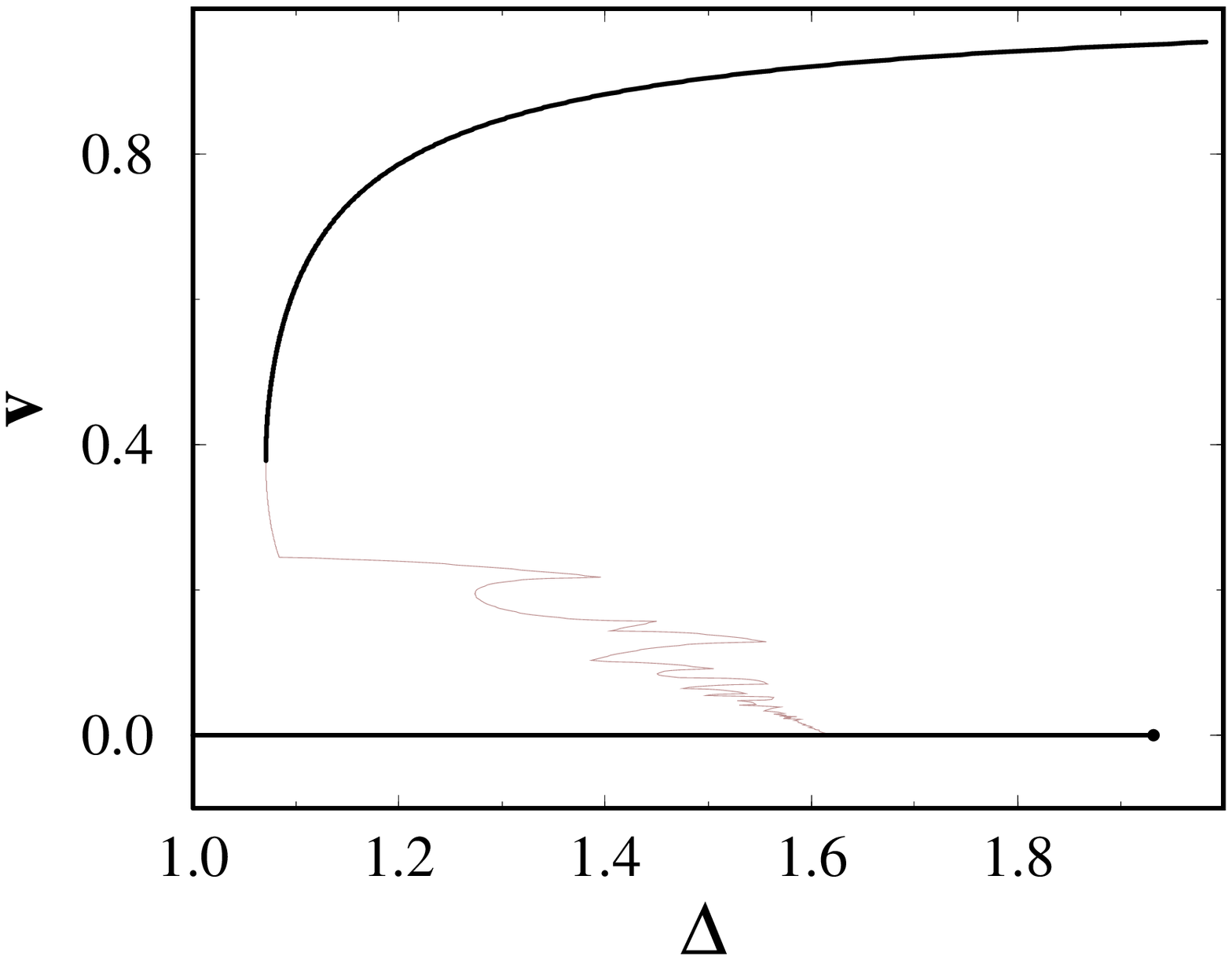}}
\endfigure

It is interesting to plot the
function $\Delta(v)$ obtained from \Eq{OD20}
(\Fig{oned.energy}). Because all steady states occur for $\Delta>1$, one
necessarily concludes that not all energy stored to the right of the
crack tip ends up devoted to snapping bonds. The
fate of the remaining energy depends upon the amount of dissipation
$b$, and the distance from the crack tip one inspects. In the limit of
vanishing dissipation $b$, traveling waves leave the crack tip and
carry energy off in its wake; the amount of energy they contain
becomes independent of $b$. Such a state is depicted in \Fig{balls},
which shows a solution of \Eq{L93.6} for $v=0.5$, $N=9$, and
$b=0.01$. For all nonzero $b$, these traveling waves will
eventually decay, and the extra energy will have been absorbed by
dissipation, but the value of $b$ determines whether one views the
process  as microscopic or macroscopic.
\figure{oned.unphysical}
\infiglist{Unphysical steady state solution.}
\Caption The height of $u_{0+}$ is plotted as a function of time for
$v=.3$, $b=.01$, and $N=9$. Notice that $u$ rises above $u_f=1$ before
$t=0$, and is actually descending at the moment when it again passes
height $u_f$ and is supposed to have broken. 
\endCaption
\epsfysize=3.5 truein
\centerline{\epsffile{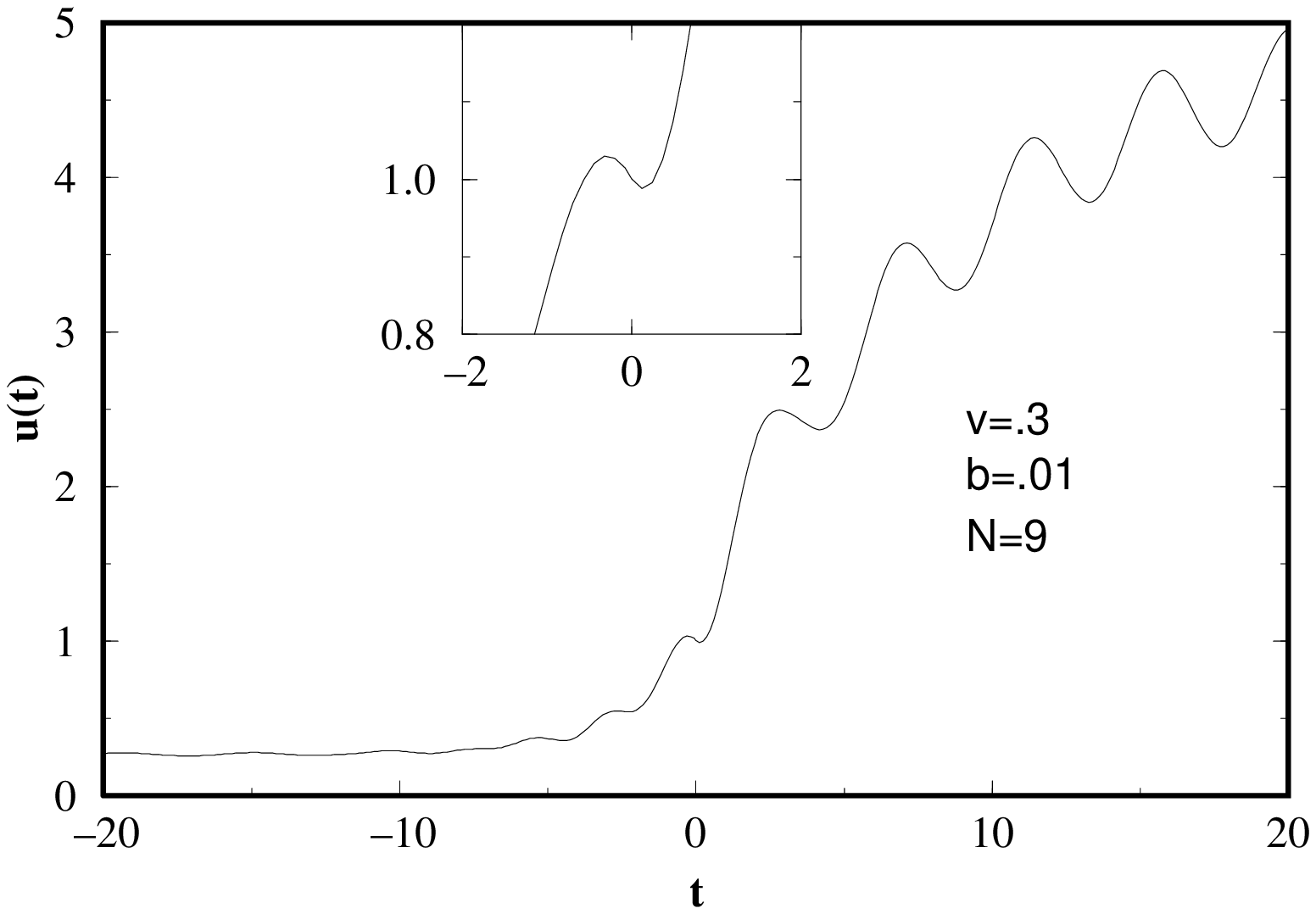}}
\endfigure

\subsection{Forbidden velocities}

The jagged structure of \Fig{oned.energy} makes it appear that many
different states, emitting different quantities of radiation, can
coexist for some values of $\Delta$. This conclusion is largely incorrect, for
two reasons. In Appendix III, it is shown that states are linearly
unstable whenever $v$ is a decreasing function of $\Delta$. So all the
backward-traveling portions of the curve can be ruled out. In addition
a final condition has been neglected. Not only must the bond between
$u_{0+}$ and $u_{0-}$ reach length $u_f$ at $t=0$, but this must be the
first time at which that bond stretches to a length greater than $u_f$.
For $0<v<0.3\dots$ (the precise value of the upper limit varies with
$b$ and $N$) that condition is violated. The states have the
unphysical character shown in \Fig{oned.unphysical}. Masses rise above
height $u_f$ for $t$ less than $0$, the bond connecting them to the
lower line of masses remaining however intact, and then they descend,
whereupon the 
bond snaps. Since the  solutions of \Eq{OD11} is unique, but does not
in this case solve \Eq{OD10}, no
solutions of \Eq{OD10} exist at all at these velocities. There is a
forbidden band of velocities.

Nevertheless, multiple solutions for some values of $\Delta$ are still
possible. The phenomenon of lattice trapping \rr{Thomson}
{Thomson, 1986}
 allows a crack
to sit still in a lattice under some range of external strains, before
the first bond holding it snaps and it begins to extend rapidly. The
lattice trapped solutions of this model are constructed in Appendix I,
and shown to exist in the range 
$${\sqrt 3-1\over \sqrt 2}=.5176\dots<\Delta<{\sqrt 3 +1\over \sqrt
  2}=1.931\dots .\EQN LTRAP$$
These bounds do not correspond  to the value of $\Delta$
obtained from \Eq{OD20} as $v\rightarrow 0$; that limit is carried out
in Appendix II, and shown to be $\Delta=(\sqrt 5+1)/2$. 

\subsection{Linear Stability\label{sec.lstb}}

The stability of steady states can be studied in a straightforward
manner, by adding a small term $u^1$ to them and
linearizing in the perturbation. One finds that even for stable
states, transients 
converge slowly to the final state, at rate $e^{-bt}$, where $b$ is
the damping in the model.

This final result can be established in a simple way which relies only upon
time reversal symmetries as follows:

Consider any equation of motion for some variables $u_m(t)$ of the
form
$$\ddot u_m(t)={\cal O}_m(\vec u)-b{\dot  u}_m(t),\EQN LS1.$$
where ${\cal O}$ is  invariant under time translation, and even under
time reversal. If one starts with a base solution $u^0(t-m/v)$, then
perturbations of the form
$e^{qt}u_m^1(t)=e^{qt}u^1(t-m/v)$ will obey the eigenvalue equation
$$q^2u_m^1+2q\dot u_m^1+\ddot u_m^1=\para{{\cal O}_m},\vec u,\cdot\vec
u^1-b{\dot   u_m}^1\EQN LS.2$$
One easily checks that $\dot u^0(t-m/v)$  is an eigenfunction with
eigenvalue $q=0$. In addition notice that 
$$u_m^1=\dot u^0(-t-m/v)\EQN LS.3$$
 obeys
$$\ddot u_m^1=\para{{\cal O}_m},\vec u,\cdot\vec u^1+b{\dot
  u}_m^1\EQN LS.4;a$$
$$\Rightarrow -2b \dot u_m^1+\ddot u_m^1=\para{{\cal O}_m},\vec
u,\cdot \vec u^1-b{\dot 
  u_m}^1.\EQN LS.4;b$$
Comparing with \Eq{LS.2}, one sees that  to first order in $b$, one
has an eigenfunction with eigenvalue 
$-b$, given simply by time reversing $\dot u^0$.

This general conclusion is reproduced by the much more detailed
analysis of Appendix IV, as discussed in Section \use{lstb.mode.III}.

\section{Simple Two-dimensional Model (Mode III)\label{sec.modeIII}}
\subsection{Definition and Energetics of the Model}
\figure{93.1}
\infiglist{Drawing of Lattice Model}
\Caption
Lattice model of fracture. The equilibrium
locations of  mass points are indicated by the white dots, while
the black dots indicate the displacements $u(m,n)$  of mass points out
of the page once stress is applied. The top line of dots is displaced
out of the page by amount $u(m,N+1/2)=\Delta\sqrt{2N+1}$, and the bottom line
into the page by amount $u(m,-N-1/2)=-\Delta\sqrt{2N+1}$. Lines connecting mass
points indicate whether the displacement between them has exceeded the
critical value of $2$ (see \Eq{L93.1;b}) . The crack tip has just
reached location $m=0$.  
\endCaption
\epsfxsize=6 truein
\centerline{\epsffile{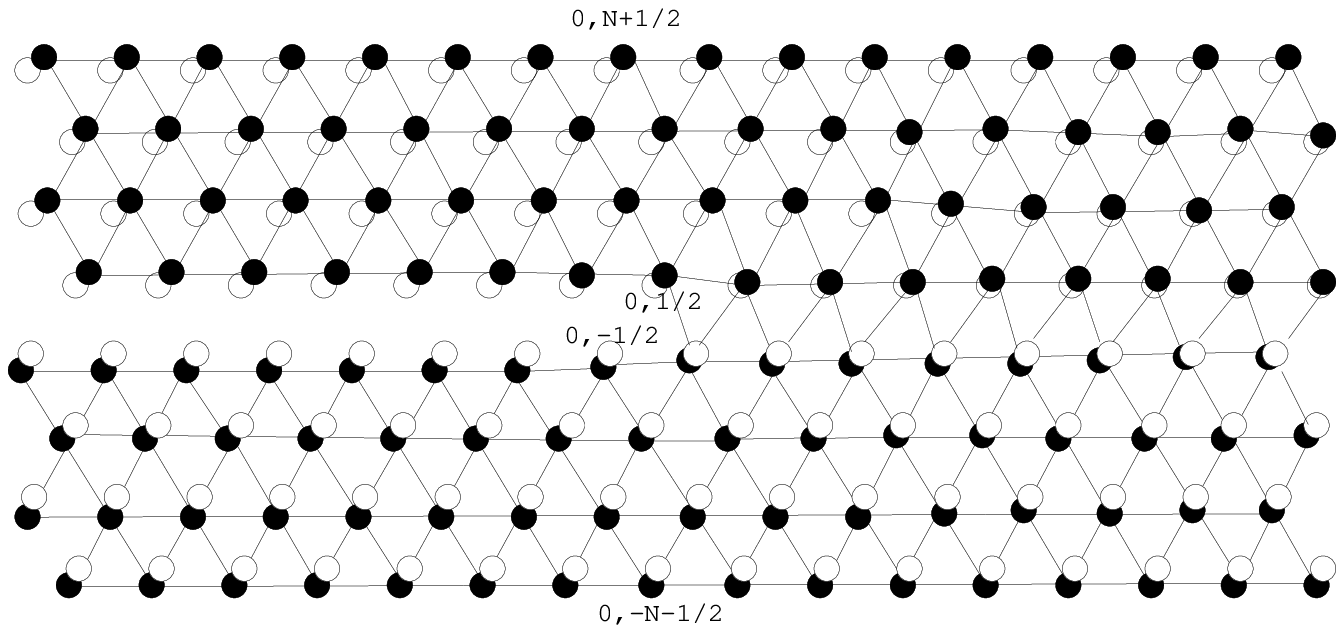}}
\endfigure

The calculation of the previous section will now be extended to a
two-dimensional lattice model, depicted in \Fig{93.1}. 
A crack moves in a
lattice strip composed of $2(N+1)$ rows of mass points. All of the bonds
between lattice points are brittle-elastic, behaving as perfect linear
springs until the instant they snap, from which  point on they exert no
force. The location of each mass point is
described by a single spatial coordinate $u(m,n)$, which can be
interpreted as the height of mass point $(m,n)$ into or out of  the
page. The index $m$ takes integer values, while $n$ takes
values of the form $1/2$, $3/2$, $\dots N+1/2$.
The model is described by the equation
$$
{\ddot u}(m,n)= -b\dot u+{1\over 2} \sum_{{\rm nearest\atop neighbors}
m^\prime,n^\prime} \F[u(m^\prime,n^\prime)-u(m,n)],\EQN L93.1;a$$
with
$$\F(u)=u\theta(2u_f-|u|)\EQN L93.1;b$$
representing the brittle nature of the springs,  $\theta$ the step
function, and $b$ the coefficient of a small dissipative
term. The boundary condition which drives the motion of the crack is 
$$u(m,\pm[N+1/2])=\pm U_N\EQN L93.1;c$$

As before, it is important to find the value of $U_N$ for which there
is just enough energy stored per length to the right of the crack to
snap the pair of bonds connected to each lattice site on the crack line.
For $m\gg 0$ one has
$$u(m,n)={n U_N \over N+1/2}\EQN L93.1.1$$
so that the energy stored per length in the $2N+1$ rows of bonds is
$${\rm {1\over 2}\times[2\ Upper\ Bonds/Site]\times[Rows]\times[Spring\
Constant]\times} U_{\rm right}^2$$
$$ ={1\over 2} 2(2N+1){1\over 2} ({U_N\over N+1/2})^2\EQN L92.1.2$$
$$=2Q_0(U_N)^2,\EQN L93.1.3$$
with $Q_0=1/(2N+1)$ given as before by \Eq{OD3}.
The energy required to snap two bonds each time the crack advances by
unit length is 
$${1\over 2}{1\over 2}(2u_f)^2+{1\over 2}{1\over
2}(2u_f)^2=2u_f^2.\EQN L93.1.4$$ 
Therefore, as before, the proper measure of external driving is
$$\Delta={U_N\sqrt{Q_0}\over u_f},\EQN L93.1.4.1$$
a quantity which reaches $1$ as soon as there is enough energy to the
right of the crack to snap the bonds along the crack line.

\subsection{Reduction to Form of Previous Section}
The techniques used to solve this model were found some time ago by
Slepyan \rr{Slepyan}
{Slepyan, 1981}. There are differences between details of his
solution and ours
because \Eq{L93.1}
describes motion in a strip rather than an infinite plate, and in a
triangular rather than a square lattice. The
strip is preferable to the infinite plate when it comes time to
compare with numerical simulations, while reducing to the simpler
infinite plate results in certain natural limits.

Assume that a crack moves in this lattice in steady state, so that
one by one, the bonds connecting 
$u(m,1/2)$ with  $u(m+1,-1/2)$ or $u(m,-1/2)$ break. They break
because the distance between these points exceeds the limit set in
\Eq{L93.1;b} and as a consequence of the driving force described by
\Eq{L93.1;c}. Supposing that these are the only bonds which snap (an
assumption to which we will return later) it is possible to calculate the
motion of all the mass points above the line $n=1/2$ as a function of the
mass points on the line $n=1/2$, since in any region where the bonds
do not snap the model has simple traveling wave solutions. 

In steady state, one has the symmetry
$$u(m,n,t)=u(m+1,n,t+1/v)\EQN TL3;a$$
and also
$$u(m,n,t)=-u(m,-n,t-[1/2-g_n]/v)\EQN TL3;b$$
which implies in particular that
$$u(m,1/2,t)=-u(m,-1/2,t-1/2v).\EQN TL3;c$$
We have defined
$$g_n=\cases{0&if $n=1/2,5/2\dots$\cr
             1&if $n=3/2,7/2\dots$\cr
             {\rm mod}(n-1/2,2)&in general},\EQN TL1$$
One can now eliminate the variable $m$ entirely from the equation of
motion, by defining 
$$u_n(t)=u(0,n,t),\EQN 93a0$$
and write the equations of motion in steady state as
$${\ddot u}_n(t)=\eqalign
{{1\over 2}&[\eqalign{+u_{n+1}(t-(g_{n+1}-1)/v)&+u_{n+1}(t-g_{n+1}/v)\cr
+u_{n}(t+1/v)-6&u_{n}(t)+u_{n}(t-1/v)\cr  
 +u_{n-1}(t-(g_{n-1}-1)/v)&+u_{n-1}(t-g_{n-1}/v) }]\cr -&b{\dot
u}_n}\EQN TL2.1;a$$  
if $n>1/2$, and
$$\eqalign{ &{\ddot u}_{1/2}(t)=\cr
&{1\over 2}[\eqalign{+u_{3/2}(t)&+u_{3/2}(t-1/v)\cr
+u_{1/2}(t+1/v)-4&u_{1/2}(t)+u_{1/2}(t-1/v)\cr  
+ [u_{-1/2}(t)-u_{1/2}(t)]\theta(-t)
&+[u_{-1/2}(t-1/v)-u_{1/2}(t)]\theta(1/(2v)-t)}]\cr
&-b{\dot u}_{1/2}}.\EQN TL2.1;b$$ 
The time at which the bond between $u(0,1/2,t)$ and $u(0,-1/2,t)$ breaks
has been chosen to be $t=0$, so that by symmetry the time the bond between
$u(0,1/2,t)$ and $u(1,-1/2,t)$ breaks is $1/2v$.

For $n>1/2$ it is easy to solve the linear set of equations \Eq{TL2.1;a}.
Fourier transforming in time gives  
$$-\omega^2 u_{n}(\omega)=ib\omega
+\eqalign{{1\over 2} u_{n+1}(\omega)&[e^{i\omega(g_{n+1}-1) /v} +
e^{i\omega(g_{n+1}) /v} ]\cr
+{1\over 2}u_{n}(\omega)&[e^{i\omega /v} -6+
e^{-i\omega/v} ]\cr
{1\over 2}u_{n-1}(\omega)&[e^{i\omega(g_{n-1}-1) /v} +
e^{i\omega(g_{n-1}) /v} ]}.\EQN TL4$$
Let
$$u_{n}(\omega)=u_{1/2}(\omega)e^{k(n-1/2)-i\omega g_n/(2v)}.\EQN TL5$$
Substituting this expression into \Eq{TL4}, and noticing that
$g_n+g_{n+1}=1$ gives
$$-\omega^2u_{1/2}(\omega)=ib\omega u_{1/2}(\omega)
+\eqalign{
{1\over 2}u_{1/2}(\omega)e^k&[e^{i\omega(g_{n+1}+g_n-2) /(2v) } + 
e^{i\omega (g_{n+1}+g_n) /(2v)} ]\cr
{1\over 2} u_{1/2}(\omega)&[e^{i\omega /v} -6+
e^{-i\omega/v} ]\cr
+{1\over 2} u_{1/2}(\omega)e^{-k}&[e^{i\omega(g_{n-1}+g_n-2) /(2v)} +
e^{i\omega(g_{n-1}+g_n) /(2v)} ]}\EQN TL6$$
$$\Rightarrow \omega^2+ib\omega+2\cosh( k)\cos(\omega/(2v))
+\cos(\omega/v)- 3=0.\EQN TL7$$
Defining
$$z={3-\cos(\omega/v)-\omega^2-ib\omega\over 2\cos(\omega/2v)}\EQN
TL8$$ 
one has equivalently that
$$y=z+\sqrt{z^2-1},\EQN TL9$$
with 
$$y=e^k.\EQN TL10$$
One can construct a solution which meets all the boundary conditions
by writing
$$u_n(\omega)= u_{1/2}(\omega)e^{-i\omega g_n/2v}
[{y^{[N+1/2-n]}-y^{-[N+1/2-n]}\over 
y^{N}-y^{-N}}]+{U_N (n-1/2)\over N}{2\alpha\over
\alpha^2+\omega^2}.\EQN TL11$$ 
This solution equals $u_{1/2}$ for $n=1/2$, and equals $U_N
2\alpha/(\alpha^2+\omega^2)$ for $n=N+1/2$, with $\alpha$ sent to zero
at the end of the calculation.
The most interesting variable is not $u_{1/2}$, but the
distance between the bonds which will actually snap. For this reason define
$$U(t)={u_{1/2}(t)-u_{-1/2}(t)\over
2}={u_{1/2}(t)+u_{1/2}(t+1/2v)\over 2}. \EQN TL12$$

Rewrite \Eq{TL2.1;b} as
$${\ddot {u_{1/2}}}(t)=\eqalign{ 
&{1\over 2}[\eqalign{+u_{3/2}(t)&+u_{3/2}(t-1/v)\cr
+u_{1/2}(t+1/v)-4& u_{1/2}(t)+u_{1/2}(t-1/v)\cr  
- 2U(t)\theta(-t)
&-2U(t-1/2v)\theta(1/(2v)-t)}]\cr
&-b{\dot {u_{1/2}}}}.\EQN TL13$$ 
Fourier transforming this expression  using \Eq{TL11} and 
defining
$$U^\pm(\omega)=\int_{-\infty}^\infty d\omega e^{i\omega t}
U(t)\theta(\pm t),\EQN L93.5.1$$  
now gives 
$$ u_{1/2}(\omega)F(\omega)-(1+e^{i\omega/2v})U^-(\omega)
=-{U_N\over N}{2\alpha\over\omega^2+\alpha^2},\EQN TL14$$
with
$$F(\omega)=\lb{{y^{[N-1]}-y^{-[N-1]}\over
y^{N}-y^{-N}}}-2z\rb\cos(\omega/2v)+1\EQN L93.5.1b$$
Next, use \Eq{TL12} in the form
$$U(\omega)={(1+e^{-i\omega/2v})\over 2} u_{1/2}(\omega)\EQN TL15$$
to obtain
$$ U(\omega)F(\omega)-2(\cos^2{\omega/4v})u^-(\omega)
=-{U_N\over N}{2\alpha\over\omega^2+\alpha^2}.\EQN TL14$$
Writing
$$U(\omega)=U^+(\omega)+U^-(\omega) \EQN L93a1$$
finally gives
$$ U^+(\omega) Q(\omega) + U^-(\omega)
={U_N Q_0} 
\lb{1\over
  \alpha+i\omega} +{1\over\alpha-i\omega}\rb
,\EQN L93.5$$
with 
$$Q(\omega)={F(\omega)\over F(\omega)-1-\cos (\omega/2v)}.\EQN
L93.5.1$$
To obtain the right hand side of \Eq{L93.5} one uses the facts that
$F(0)=-1/N$, and that $\alpha$ is very small, so that the right hand side of
\Eq{L93.5} is a delta function. 

Notice as promised that \Eq{L93.5} is identical to \Eq{OD16}.
Therefore from this point forward the analysis of the Section \use{sec.1D}
can be repeated without alteration. In particular \Eq{L93.6} gives an
explicit expression for $U(\omega)$, and \Eq{TL26} and \Eq{OD20}
describe the relation between $\Delta$ and $v$.

Without repeating the calculations, we mention for later reference
that a square lattice can be solved in the same manner -- it is
simpler than the triangular lattice -- and the results are the same
except that
$$F(\omega)={{y^{[N-1]}-y^{-[N-1]}\over
y^{N}-y^{-N}}}-2z+1\EQN CL20.0$$
replaces \Eq{L93.5.1b} and
$$Q={F\over F-2}\EQN CL20.1$$
replaces \Eq{L93.5.1}.

\figure{uout}
\infiglist{}
\Caption
Plot of $U(t)$ for $v=.5$, $N=9$, and $b=0.01$, produced by direct
evaluation of \Eq{L93.6}. Note that mass points are nearly motionless until
just before the crack arrives, and that they oscillate afterwards for
a time on the order of $1/b$.
\endCaption
\epsfysize=4 truein
\hskip 1.5truein\epsffile{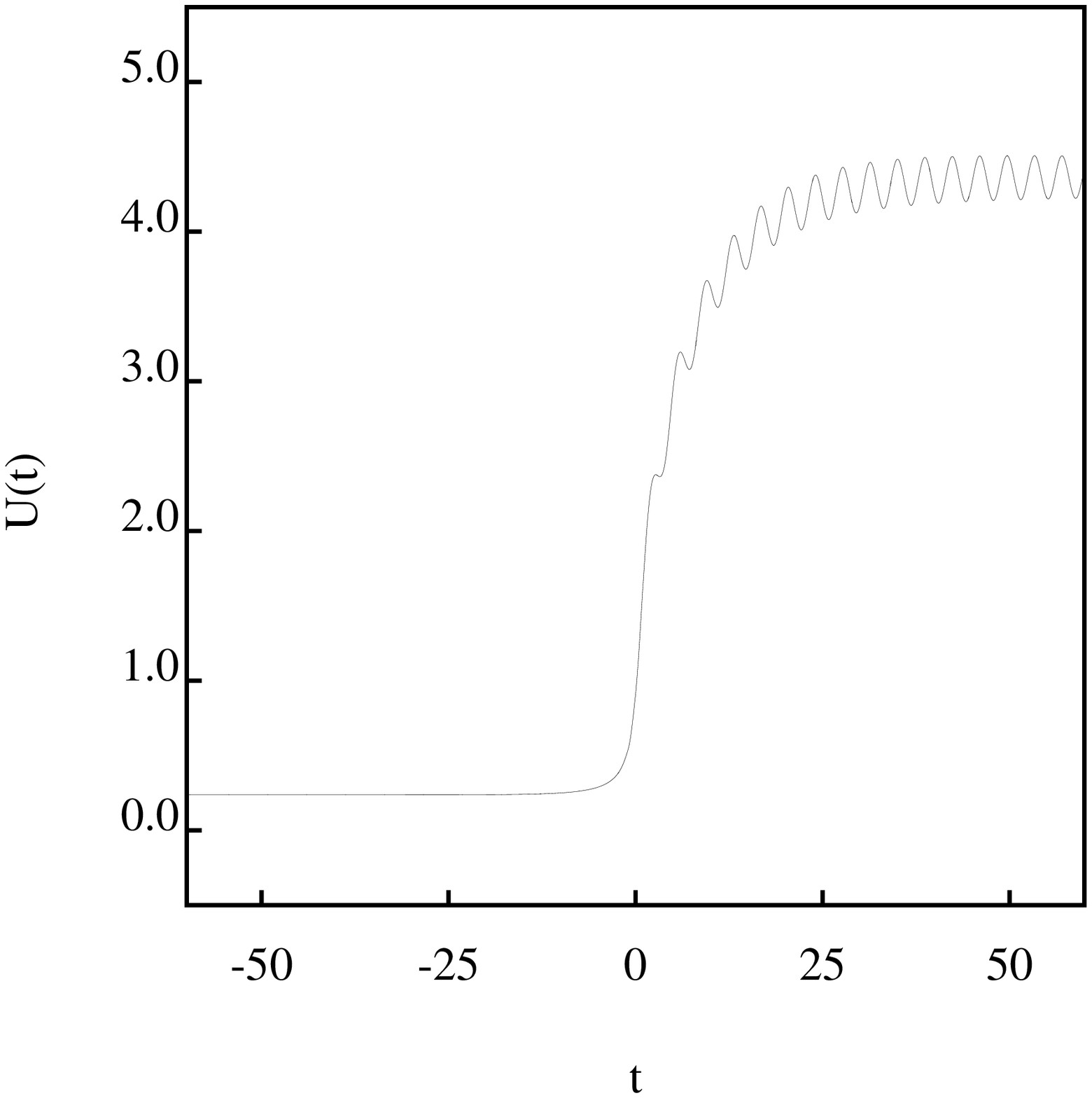}
\endfigure

There are three parameters to vary in looking for numerical solutions
of \Eq{L93.5}.
The most important is $v$, the velocity of the steady state. In
addition, one can also control $N$, the width of the strip, and $b$
the amount of dissipation. There is a  natural limit in which many results
become independent of $N$ and $b$, the limit of a macroscopic
dissipationless strip,  namely  $N\rightarrow\infty$ and
$b\rightarrow 0$. No physical results depend upon the order in which
these limits are taken, although the integrals one has to perform for
\Eq{TL26} look very different. For $1/b\gg N$, $\ln{Q/\bar Q}$ is only
nonzero near the finite number of points where $Q$ has a pole or a
zero. One can use \Eq{OD20} in this case. For $N\gg 1/b$ these poles and
roots merge into branch cuts,  
and the integrand of \Eq{TL26} becomes smooth. By taking the limit in
this way, one can show the equivalence of results in the strip with
the results found previously by Slepyan \rr{Slepyan}
{Slepyan, 1981} for a square
lattice occupying an infinite plate. In the
following discussion, the calculations will be carried out for $N=9$.
Although seemingly far from the limit $N\rightarrow \infty$ , all
physical quantities that 
have been checked so far change only in the fourth decimal place when
$N$ increases from $9$ to $100$. So $N=9$ is large enough  to give an
accurate picture of the behavior of the model, but small enough to
make all types of computations rapid. 

All the phenomena discussed in relation to the one-dimensional model
occur here. A picture of $u(t)$ for $v=.5$ appears in \Fig{uout}, a
graph of $\Delta(v)$ appears in \Fig{93.2},  and a picture of an
unphysical state at $v=.2$ appears in \Fig{u.20.05}.
For $N=9$
and $b=0.05$, starting at the wave speed $c=\sqrt{3}/2$ and working
downwards, $\dot 
u(0)$ first becomes negative at $v=0.244$, or 28\% of the wave speed $c$.   At lower
values of $v$, there is no evidence that the steady states are ever physical.
For example, at $v=0.106$, $\dot u(0)$ is positive. However,
examining the state, one  finds that $u$ earlier rose above $u(0)$,
descended below it, and is on the rise again. Stationary states with $v=0$ are
physical, and exist up until the point where they are met by the
states with $v\neq 0$. Once again, there is a velocity gap, and no
steady states exist 
between $v=0$ and $v=0.244\dots$. Above $v=0.577$, or $v/c=.666$,  the
steady state solutions are unstable, for reasons that will be
discussed in a later section.

\figure{93.2}
\infiglist{Delta versus v}
\Caption
Steady state velocities as a function of external
strain $\Delta$, calculated for strip half-width $N=9$ and dissipation
$b=0$ using \Eq{OD20}.  The thick lines indicate cases in which the steady 
states are stable. Zero velocity states at strains
$\Delta>1$ correspond to the phenomenon of lattice trapping
\rr{Thomson}
{Thomson, 1986}.
\endCaption
\epsfysize=3 truein
\centerline{\epsffile{mode3.roots.9.vgrps}}
\endfigure

\figure{u.20.05}
\infiglist{Behavior of $U(t)$ for $v=.2$, $b=0.05$, and $N=9$.}
\Caption
Behavior of $u(t)$ for $v=.2$, $b=0.05$, and $N=9$. Notice that at
$t=0$, indicated by the dashed line,  $u$ is decreasing, and that it
had already reached height $1$  earlier. This state is not physical.
\endCaption
\vskip 3.5 truein
\includegraphics{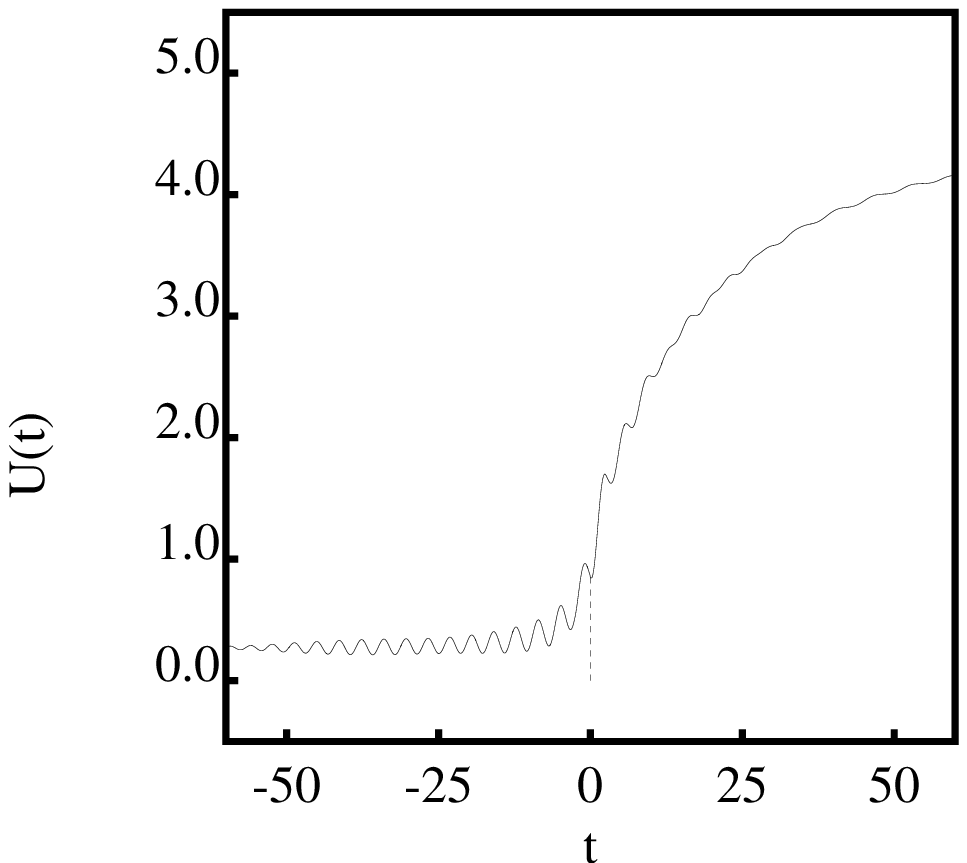}
\endfigure

\subsection{Linear Stability \label{lstb.mode.III}}

The complete calculation of linear stability is carried out for the
triangular lattice of this section and the details are
relegated to Appendix IV. The results are as follows:

Define
$$z_q={3-\cos(\omega/v)+(q-i\omega)^2-ib\omega\over 2\cos(\omega/2v)}\EQN
LP3;a$$ 
$$y_q=z_q+\sqrt{z_q^2-1},\EQN LP3;b$$
and
$$F_q(\omega)=\lb{ y_q^{[N-1]}-y_q^{-[N-1]}\over
y_q^{N}-y_q^{-N}}-2z_q\rb \cos(\omega/2v)+1.\EQN LP4.1$$
Define in addition
$$U(t)={u^1_{1/2}(t)-u^1_{-1/2}(t+1/2v)\over 2}\EQN LP5;a$$
and restrict attention to  modes with the symmetry 
$$u^1_{1/2}(t)=-u^1_{-1/2}(t+1/2v),\EQN LP5;b$$
Then defining
$$Q_q(\omega)={F_q(\omega)\over F_q(\omega)-1-\cos (\omega/2v)}\EQN L93a.8$$
one has
$$U(\omega)=(Q_q^+-{Q_q^-}){U_0\over
\dot u^0(0)} ,\EQN L93a.6$$
with $\dot u^0(0)$ the velocity of the steady state $u^0(t)$ at $t=0$,
and $U_0$
subject to the consistency condition
$$U_0=\int {d\omega\over 2\pi} U(\omega),\EQN L93a.7$$
which may be rewritten as
$$\int {d\omega\over 2\pi} 
(1-Q_q^-(\omega))\equiv S_q=\dot u^0(0)=S_0. \EQN L93a.7.1$$ 
The spectrum of perturbations about steady states is given by the
zeroes of $S_q-S_0$. If $S_q-S_0$ has a root when the real part of $q$ is
positive, the corresponding steady state is unstable.

One consequence of \Eq{L93a.7.1} is that any state whose velocity decreases as
$\Delta$ increases must be unstable. This result may be established by
considering the behavior of 
$S_q$ for small $q$. If the slope of $S_q$ is positive at the origin, then
there must be a root for positive $q$, since $S_0>0$, and $S$ goes to zero as
$q\rightarrow \infty$. The slope $S^\prime_0$ is given by the
following expression, evaluated for the steady state:
$$S^\prime_0=-v\para ,v, \ln Q_0^-(0)
. \EQN LP25.01 $$
or using \Eq{L93.6}, 
$$S^\prime_0=-\para \ln\Delta,\ln v, \EQN LP26$$

When accurate numerical evaluation of $S(q)$ is necessary, it can be
carried out from the integral
$$S_q= \int{d\omega^\prime \over 2\pi} \ln [{Q_0(\omega^\prime)\over
Q_q(\omega^\prime)}] .\EQN LP30$$

The moderately lengthy derivations of \Eq{LP26} and \Eq{LP30} are
given in Appendix V. 
\figure{perta}
\infiglist{}
\Caption
The integral on the right side of \Eq{LP12} is plotted as a function
of $q$ for $v=0.5$ and $b=0.0125$. 
\endCaption
\vskip 3.5 truein
\includegraphics{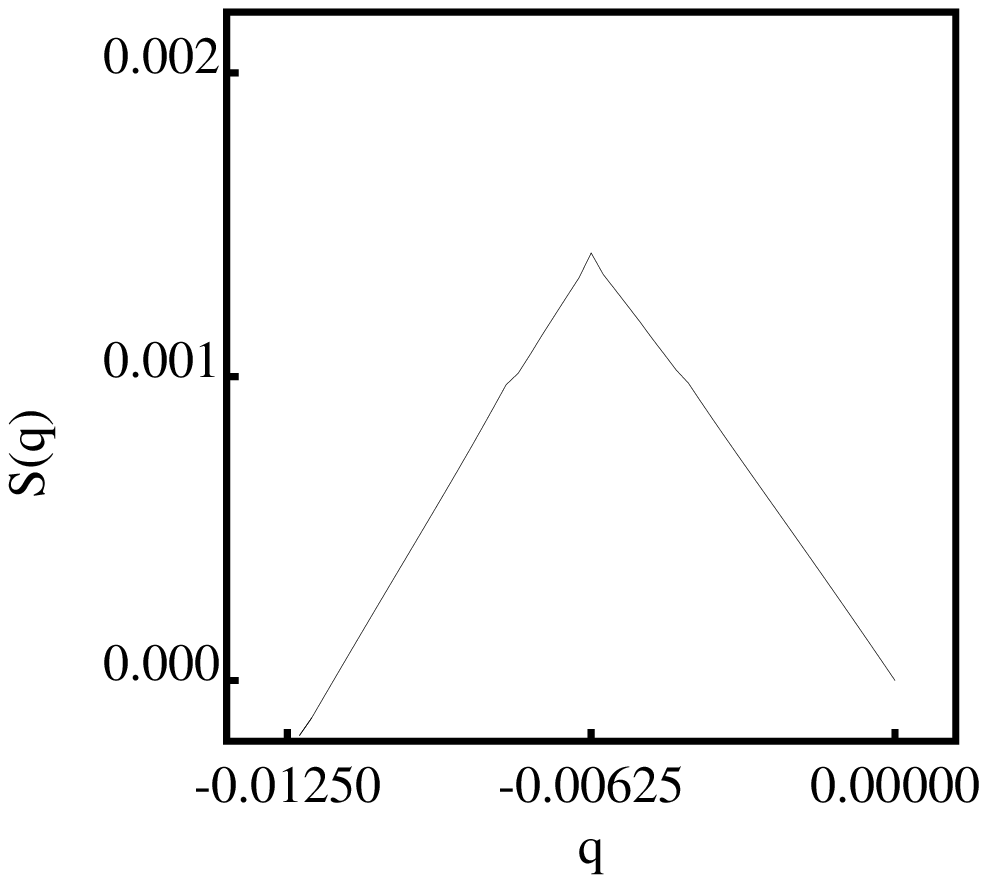}
\endfigure

\Fig{perta} shows a graph of $S_q$. The somewhat surprising triangular
shape has a clear physical origin. According to \Eq{LP3;a}, when $q$
is real and very small, it acts so as to shift the effective value of
the damping $b$. Right at $q=-b/2$, the system is fooled into thinking
that $b$ has vanished, and as $q$ decreases further, it is as if $b$
has changed sign, or as if time has started to run backwards.
The symmetry of $Q$ implies that
$$Q(\omega,b)=Q(-\omega,-b)\EQN L93a.10 $$
$$\Rightarrow {Q^-(\omega,b) Q^+(-\omega,-b)}=
{Q^-(-\omega,-b) Q^+( \omega, b)}\EQN L93a.11$$
Since the left side of this equation has roots only above the real
axis, and the right side only below,
$$Q^-(0,-b)=1/Q^+(0,b),\EQN LP25.1$$
so using
$$u^0(0)=1=\sqrt{2N+1}{\Delta\over Q^+(0)},\EQN LP25.2$$
one has that for small positive $\epsilon$
$$S^\prime_{b/2+\epsilon}= v\para ,v, \ln Q_0^+(0)=\para \ln
\Delta,\ln v, .\EQN LP25.3$$
In short, the sign of the slope of $S_q$ changes sign right at $q=-b/2$,
resulting in a stable eigenmode with eigenvalue exactly $-b$. This
result is exactly that obtained in Section \use{sec.lstb}.  A
picture of the mode is given in \Fig{u0dot}.

\figure{u0dot}
\infiglist{}
\Caption
The eigenmode $U(t)$ for $q=b$, given by $\dot u^0(-t)$ is pictured for
$v=0.5$ and $b=0.0025$. 
\endCaption
\epsfysize=4 truein
\centerline{\epsffile{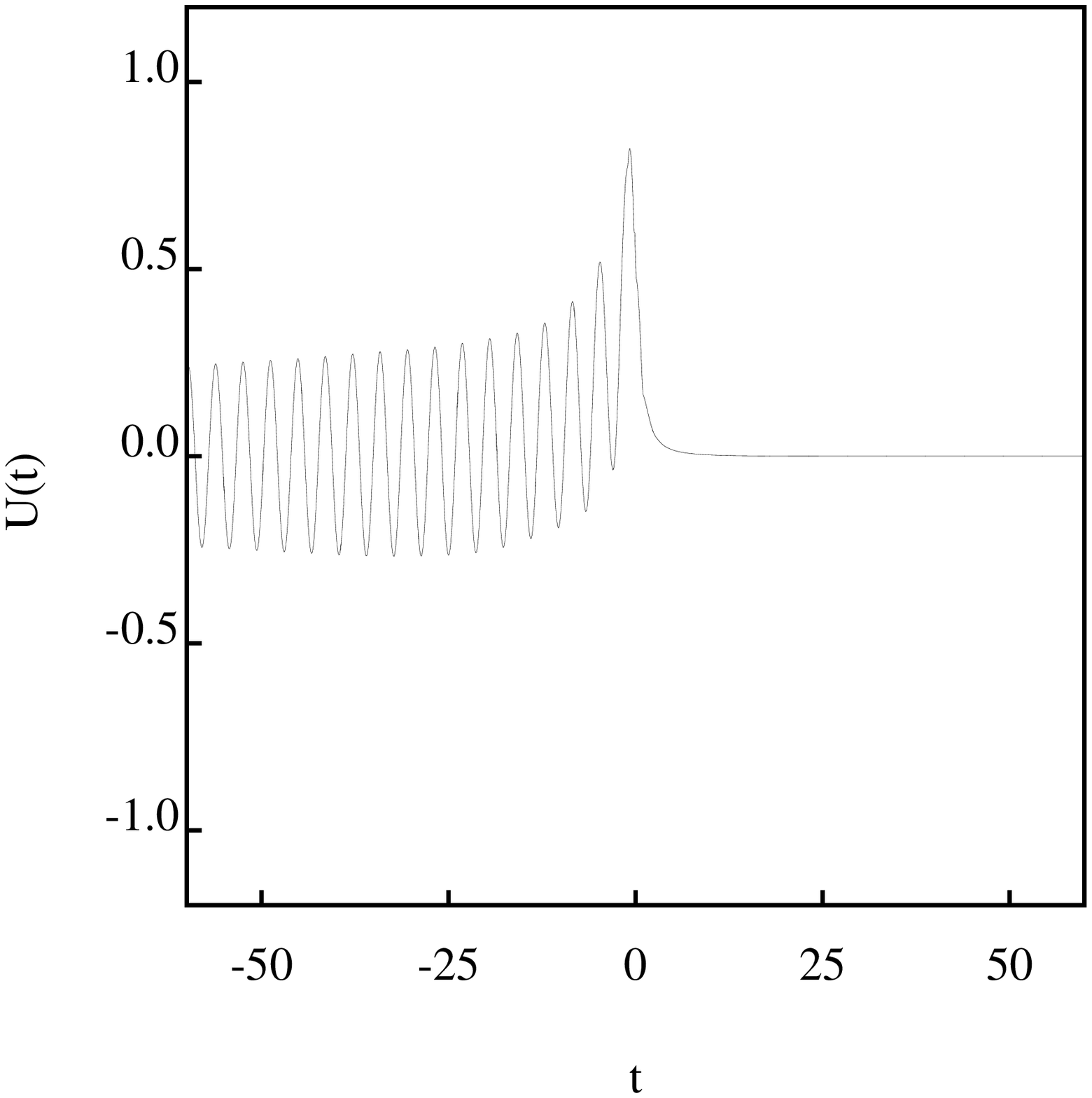}}
\endfigure

\subsection{Nonlinear Instabilities}
Showing that steady states are linearly stable is not sufficient
to demonstrate that they are physical.
It was assumed in their derivation that the only
bonds which break are those which lie on the crack path. From the
numerical solutions of \Eq{L93.6},  one
can test this assumption; it fails above a critical velocity.  
Recall that the sound speed $c$ equals $\sqrt{3}/2$. For $N=9$, at a
velocity of $v_c/c=.666\dots$, $\Delta_c=1.158\dots$, 
the bond between $u(0,1/2)$ and $u(1,1/2)$  reaches a
distance of $2 u_f$ some short time after the bond between $u(0,1/2)$ and
$u(0,-1/2)$ snaps. The steady state solutions strained with larger
values of $\Delta$ are inconsistent; only dynamical solutions more
complicated than steady states, involving the breaking of bonds off
the crack path, are possible. To investigate these states, one must  return
to \Eq{L93.1} and numerically solve the model directly. These
simulations have been carried out by  Liu \rra{Marder.PRL.93}
{Marder and Liu, 1993}
\rrc{Liu_93}
{Liu, 1993}
,
and some results are contained in \Fig{xiang}.

\figure{xiang}
\infiglist{}
\Caption
 Pictures of broken bonds left behind the crack tip
at four different values of $\Delta$, from simulations of Liu
\rr{1993}. The top figure shows the simple 
pattern of bonds broken by a steady-state crack. At a value of
$\Delta$ slightly above the critical one where horizontal bonds
occasionally snap, the pattern is periodic. All velocities are
measured relative to the sound speed $c=\sqrt{3}/2$. Notice that the average
velocity can decrease relative to the steady state, although the external
strain has increased. As the strain $\Delta$ increases further, other
periodic states can be found, and finally states with complicated
spatial structure. The simulations are carried out in a strip with
half-width $N=9$, of length 200 and $b=0.01$. The front and back ends
of the strip have short energy-absorbing regions to damp traveling waves.
\endCaption
\epsfysize=3.5 truein
\centerline{\epsffile{broken.bonds.idr}}
\endfigure

The diagram shows  patterns of broken bonds left behind the
crack tip. Just above the threshold at which horizontal bonds begin to
break, one expects the 
distance between these extra broken bonds to diverge. The reason is
that breaking a horizontal bond takes energy from the crack and
slows it below the critical value. The crack then tries once more to
reach the steady state, and only in the last stages of the approach
does another horizontal bond snap, beginning the process again.
This scenario for instability is similar to that known as
intermittency in the 
general framework of nonlinear dynamics \rr{Manneville}
{Manneville, 1990}
; the system
spends most of its 
time trying to reach a fixed point which the motion of a control
parameter has caused to disappear.

Here is a rough estimate of the distance between broken horizontal bonds.
Let $u_h(t)$ be the length of an endangered horizontal bond as a function of
time. Actually, one needs to view matters  in a reference frame moving with
the crack tip, so every time interval $1/v$, one shifts attention to
a bond one lattice spacing to the right. When $\Delta$ is only
slightly greater than $\Delta_c$, the 
length of such a bond viewed in a moving frame  should behave before 
it snaps, as 
$$u_h\sim 2  u_c+ \para u_h,\Delta,(\Delta-\Delta_c)-\delta u
e^{-bt}.\EQN GE1$$ 
Here $\partial u_h/\partial \Delta$ means that one should calculate
the rate at which the steady-state length of $u_h$ would change with
$\Delta$ if this bond were {\it not} allowed to snap, and $\delta u$
describes how much smaller than its steady-state value the bond ends
up after the snapping event occurs. From this expression, one can
estimate the time between snapping events by setting $u_h$ to $2 u_f$
and solving \Eq{GE1} for $t$. The result is that
the frequency $\nu$ with which horizontal bonds snap should scale
above the critical strain $\Delta_c$ as
$$\nu\sim -b/\ln({1\over \delta u}\para
u_h,\Delta,[\Delta-\Delta_c]),\EQN L93.11$$ 
a result that is consistent with the numerics, but hard to check
conclusively. One can calculate numerically that  $\partial u_h/\partial
\Delta= 5.5$ for the conditions of \Fig{xiang}, but $\delta u$ is hard
to find independently. Assuming that \Eq{L93.11} is correct, one finds
from the second picture of \Fig{xiang} that $\delta u=0.04$.
Further increasing the external strain $\Delta$ makes a  wide variety
of complicated behavior possible, including dendritic patterns, in the
lowest panel of \Fig{xiang} that we will see are reminiscent of experiment.

\section{Solution of Fully Two-Dimensional Model (Mode I)}\label{sec.modeI}
\def\kp{k_\|}\def\kr{k_\bot}

\figure{triang}
\infiglist{Diagram of Mode I Triangular Lattice}
\Caption
Diagram of lattice used in this section
\endCaption
\epsfysize=2.5 truein
\centerline{\epsffile{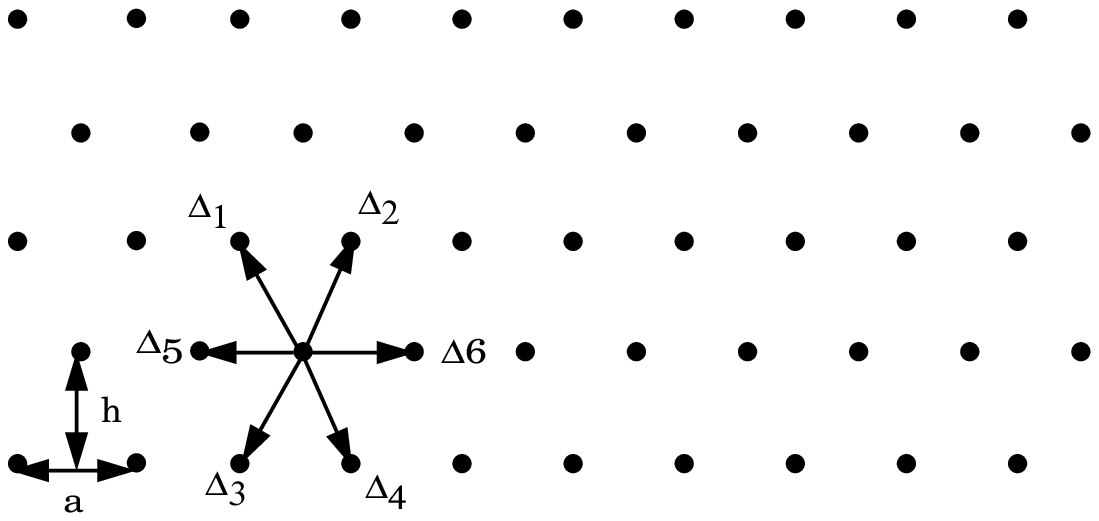}}
\endfigure

The triangular lattice solved in Section \use{sec.modeIII}  contains
all the physical 
phenomena we now know to be associated with lattice fracture. Still, it is
not very realistic because of the restriction that mass points move in
only one dimension. That restriction is relaxed in this section.
Unfortunately, the calculations rapidly become so lengthy that they
are difficult to explain. For this reason, most of the details are
contained in a MAXIMA script which was used to produce the results.
However, one can at least study the question of energy balance without
too much trouble.

The lattice studied in this section is depicted in \Fig{triang}. Each
point has two degrees of freedom associated with motion in the plane.
In the absence of external forces, the mass points all sit on a
triangular lattice, and they interact linearly with their nearest
neighbors. The interaction between two neighbors is a general linear
function of both the parallel distance between them, and the
perpendicular distance between them. The two spring constants $\kr$
and $\kp$ are introduced so as to accommodate a general Poisson ratio.
Choosing an arbitrary $\kp$ and $\kr$ results in long-wavelength
transverse and longitudinal sound speeds
$$ c_l^2={3a^2\over 8m}(\kr+3\kp)\EQN TR8;a$$
$$c_t^2={3a^2\over 8m}(3\kr+\kp),\EQN TR8;b$$
with $m$ the mass of the points, and $a$ the lattice spacing.

Unfortunately, for purely technical reasons, it is not possible to
adopt this form of the nearest neighbor force everywhere on the
lattice. In order to solve this model as in the previous
sections, it is necessary that mass points on opposite sides of the
crack experience a force only along
the line between their centers. As partial compensation, however, one
can take this spring constant $\kp^I$ to be different from the
parallel force constant $\kp$ elsewhere in the lattice. Numerical
calculations show that varying $\kp^I$ within reasonable bounds has very
little effect upon the physical results.

Taking bond interactions between neighbors in this way, one can form a
correspondence with Section \use{sec.1D}. Far ahead of the crack tip,
the ratio of vertical displacements of points right above the crack
line to vertical displacement at the top of the strip is
$$Q_0={1/ 4(\kr+3\kp)\over 1/ 4(\kr+3\kp) +2N (3/4) \kp^I}.\EQN MI.0$$
As a fracture criterion, one demands that bonds snap when the distance
between neighbors across the crack line is $2u_f$ greater than it is in
equilibrium, so that the energy needed for fracture is
$$E_{\rm break}=2 \kp^I u_f^2.\EQN MI.1$$
The displacement of the top of the strip just  necessary to supply
enough energy to snap these bonds is given again by \Eq{OD8} and
\Eq{OD8.0}. 

The main difference between the calculation here and the one in
Section \use{sec.1D} results from a consideration of the fracture
criterion. Letting $\vec u(0,1/2)$ be a point on  the crack line, its
bond with a neighbor on the other side will snap when
$$|\vec u(0,1/2)-\vec u(0,-1/2)|>2u_f\EQN MI1.1$$
$$\Rightarrow {1\over 2}[{\sqrt 3\over 2}(u^y(0,1/2)-u^y(0,-1/2))+{1\over
2}(u^x(0,1/2)-u^x(0,-1/2))]>u_f. \EQN MI.2$$
However, if one searches for the variable in terms of which \Eq{OD16}
and its sequels remains true, one finds instead
$$U={1\over 2}[(u^y(0,1/2)-u^y(0,-1/2))+{1\over
\sqrt{3}}(u^x(0,1/2)-u^x(0,-1/2))] \EQN MI.3$$
is what is needed. The reason is that far ahead of the crack, $U$
defined this way approaches $u^y$, and all the displacements far to
the right  are purely vertical. 
In terms of this variable, fracture occurs when 
$$U(t=0)=u_f{2\over\sqrt 3} ,\EQN MI.4$$
as claimed before \Eq{TL26.0}. A second difference has to do with the
form of the dissipation. Experimentally, waves of frequency $\omega$
decay at a rate $b(\omega)\sim\omega$. This does not seem well
understood theoretically, but in any event, the calculation uses
dissipation in the form
$$i\omega b_0\sqrt{\omega_0^2+\omega^2}\vec u\EQN MI.4.1$$
rather than $i\omega b u$ as before. The constant $\omega_0$ is chosen
to be something smaller than frequencies of interest.

What remains is to calculate the function $Q$ appropriate to this
lattice. It is a lengthy task, carried out with symbolic algebra, and
relegated to Appendix VI.  No really new ideas enter; some of the
results are summarized in \Fig{mode1.delta} and \Tbl{lattice.results}.
\Fig{mode1.delta} indicates that there are small separate bands of
low-velocity states, separated by small unstable regions. {\it Note,
November 1995: Mode I fracture is subject to several instabilities
beyond those studied here. The complete story has yet to be worked out.}

\figure{mode1.delta}
\infiglist{}
\Caption
Velocity $v$ measured in units of the transverse wave speed $c_t$ as a
function of $\Delta$, for the model of this section, 
calculated for $N=15, b=0.012, \kr=-.3,\kp=2.666,\kp^I=2.366$. The
thick lines are physically realizable states, the thin solid lines are
linearly unstable, and the dotted lines are unphysical states.
\endCaption
\epsfysize=3.5 truein
\centerline{\epsffile{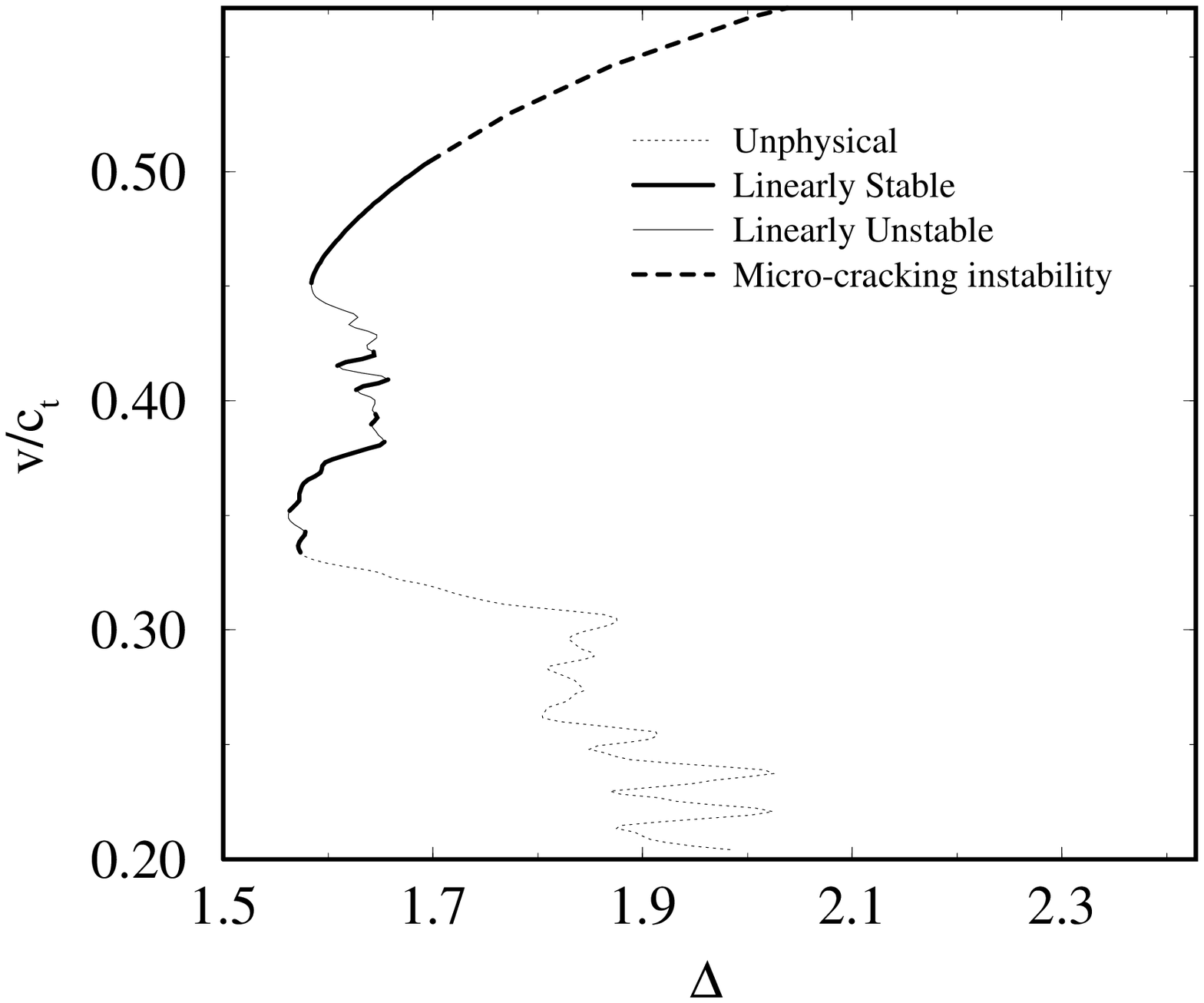}}
\endfigure

\section{Comparison with  Experiment\label{sec.exp}}

From the lattice theory emerge three predictions which can be tested
experimentally. 
\item{1.} For a band of velocities beginning at zero and proceeding up
  to around 30\% of the relevant wave speed, steady crack motion is impossible.
\item{2.} Steadily moving cracks emit radiation.
\item{3.} At around 60\% of the wave speed crack tips become unstable,
	repeatedly attempting to branch, and creating microcracks whose
	spacing is governed by the attenuation
        rate of high-frequency sound waves. The critical velocity for
	this instability depends on details of microstructure.

We have carried out experiments in Plexiglas (PMMA), differing from
those previously reported \rr{Fineberg_92}
{Fineberg, {\lowercase{\it et.~al.}}, 1991, 1992} because of improvements in
instrumentation that lead to a five-fold increase in the accuracy of
velocity measurements. The experiments are conducted in thin strips
whose geometry is chosen to match as closely as possible conditions of
the theory.
Results from the various lattice models are compared with
these experiments in \Tbl{lattice.results}. 
All of the phenomena seen in the lattice are present
in the experiment, although the velocities at which they occur are
different. Since changing from a triangular to a square lattice
changes the critical velocities substantially, this discrepancy with
experiment is not surprising. Plexiglas is a polymeric solid, and bears no
microstructural resemblance to a triangular lattice. 
It is certainly possible to study  more
complicated lattice theories numerically in search of quantitative
agreement, but in the following discussion, we want mainly to 
emphasize the qualitative correspondence between the experiments and
the theory. 

\smallskip
\table{lattice.results}
\ruledtable
Lattice | $v_{\rm min}$ | $v_c$ \cr
One-d ($N=9, b=0.01$)  |    0.3           |   -   \cr
Mode III, triangular($N=9, b=0.01$)| 0.28                | 0.67      \cr
Mode III, square($N=9, b=0.01$)| 0.36                | 0.76      \cr
Mode I ($N=15$) |0.33|0.56\cr
Mode I ($N=30$) |0.33|0.55\cr
PMMA                   | 0.2       | 0.33 
\endruledtable
\caption {Calculations of the minimum nonzero steady-state crack
velocity, $v_{\rm min}$, and the velocity $v_c$ at which the branching
instability begins for several lattices. In all cases, the
calculations are carried out in strips $2N+1$ lattice points hight.
Mode III is the lattice of 
Section \use{sec.modeIII}, and Mode I is the lattice of Section
\use{sec.modeI}. In the One-d and Mode III  cases, velocities are
measured as fractions of the sound speed, while in the Mode I case,
and in PMMA they are measured as fractions of the transverse wave
speed.  The Mode I lattice parameters are chosen to fit the
Poisson ratio, and measured ultrasonic attenuation of PMMA,
and are $\kr=-.3,\kp=2.666,\kp^I=2.366$; experimentally dissipation is
observed to be proportional to frequency, and this scaling is used in
the calculation as well, with the experimental coefficient of 0.012, 
\rr{Jackson_1972}
{Jackson, Pentecost, and Powles, 1972}
. The comparison has been made with
the plane strain  Poisson ratio; to compare with the plane stress
Poisson ratio, one can use $\kr=0$ instead of $\kr=-.3$, but the
results are not substantially different.}
\endtable
\smallskip

The first prediction has been verified whenever crack
dynamics have been measured carefully in brittle
materials \rr{Takahashi}
{Takahashi, Matsushige, and Sakurada 1984}
, although not generally
given much significance. In PMMA, the jump from quasi-static to rapid
motion goes to a velocity of around 175m/s, which is 18\% of the
Rayleigh wave speed; three of our experimental runs at three levels of
externally imposed strain are shown in \Fig{Expt1}.   The large
accelerations seen in this figure are not inconsistent with continuum
theory, as one sees by comparing with \Fig{vel}. From \Eq{L93.0.1}
with $c_l=2000 \rm m/sec$ and $w=4\ \rm cm$, 
continuum theory predicts accelerations on the order $10^8$ m/sec$^2$,
or $10^2$ m/sec/$\mu$-sec. However, it is difficult within a continuum
framework to account for the fact that cracks can be made to initiate
for such a wide range of external conditions, and that the velocity at
which rapid acceleration abruptly terminates is fairly independent of the
amount of energy pouring into the crack tip. From the perspective of
the lattice theories, both these facts are natural. Crack motion is an
activated process, and can therefore begin for a range of external
conditions, while the rapid acceleration ends when the crack velocity
has passed through the forbidden band.

Conceptually, one likes to think 
of cracks becoming unstable when their infinitesimal extension releases more
energy than it consumes, so that the crack slowly accelerates up to
high velocities. Unfortunately these slow extensions are dynamically
forbidden. The crack must begin its dynamic life at high velocities,
and the criterion for fracture initiation must be understood as a
nucleation event.  These remarks would seem to contradict the fact of
quasi-static crack motion. In the context of the lattice models, the
contradiction can be resolved if 
quasi-static growth is always a thermally or chemically aided process.
In addition, apparent quasi-static motion may actually be a stick-slip
process, in which a crack alternates between rapid motion and arrest
on scales too fast for ordinary measurements to detect.   This
observation may have implications for fracture testing. Recent
measurements of the fracture energy of PMMA have all been in the range
of 200 to 350 J/m$^2$ \rr{Katsamanis}
{Katsamanis and Delides, 1988}
. Our experiments involve long
center edge cracks in long
strips, so that the fracture energy can be deduced simply by measuring
the stress per area in the strip. Inspection of \Fig{Expt1} shows that
the fracture energy of our samples can be as low as  80 J/m$^2$ -- three
times less than typical previous measurements.{\it Note, November,
1995: These  low fracture energies have not been reproduced in
subsequent experiments, and this claim is incorrect. Fracture energies
are consistent with previous measurements.} One
gets a sense of how close one is to the minimum  fracture
energy by watching the rate of acceleration after the crack initiates.
We suspect that many of the experimental techniques used to measure
fracture energies involve dynamical effects at a more important level
than has been appreciated, but this point requires additional
investigation.

\figure{Expt1}
\infiglist{Pictures}
\Caption Three measurements of velocity versus time in Plexiglas
(PMMA). The experiments are conducted in strips, and the energy stored
per unit area to the right of the crack is printed each picture.
Cracks are made to initiate with different energies available per
length by preparing the notch which initiates fracture in different ways.
Even if the system is stressed so gently that the crack does
not accelerate noticeably after it begins rapid motion, there is still
a jump in velocity up to around 200 m/s. The 
upper part of the figure should be compared with \Fig{vel}. The
fracture energy of the sample depicted in the lower part of the figure
is several times lower than any other value for PMMA in the literature; see
Katsamanis and Delides, \rr{Katsamanis}
{1988}.  
\endCaption
\epsfysize=4truein
\centerline{\epsffile{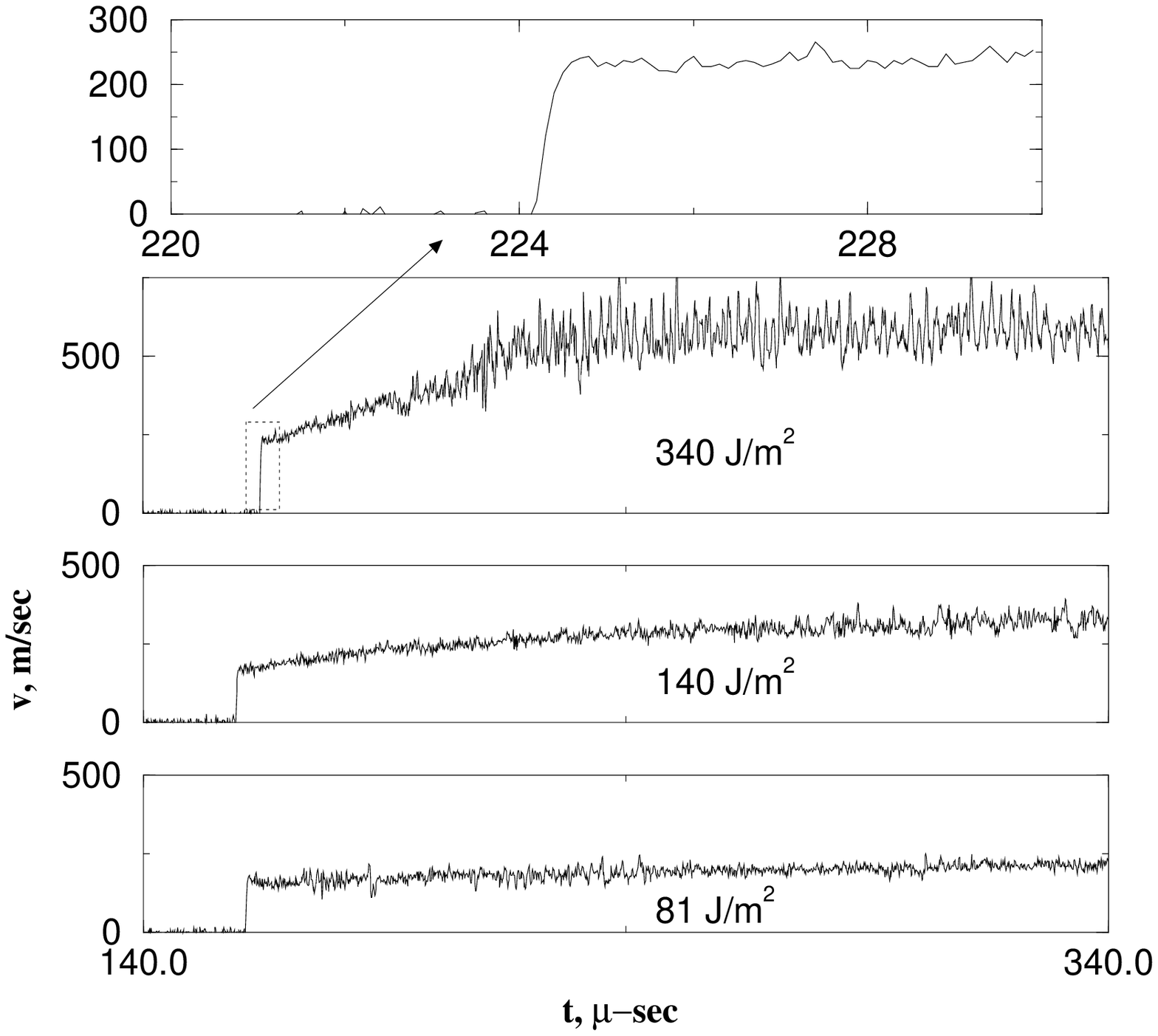}}
\endfigure

The second of the predictions has not really been verified
experimentally. It is true that moving cracks always emit
sound \rr{Gross_93}
{Gross, {\lowercase{\it et.~al.}}, 1993}.
However, acoustic transducers detect only up to around 10 MHz, which
is far below the frequencies at which atomic bonds snap when a crack
moves at hundreds or thousands of meters per second. At such high
frequencies, one should expect sound to thermalize rapidly, and
manifest itself as heat. Certainly there is always a large temperature rise
near crack tips \rra{Pratt_and_Green}
{Green and Pratt, 1974}
\rrc{Fuller}
{Fuller, Fox, and Field, 1983}
, but there are many potential
sources for it, especially plastic 
deformation. To the extent that the high-frequency sound decays within
the core region surrounding the crack tip, it causes no conceptual
difficulty for the continuum picture of fracture. Our prediction is
that the size of the heated zone should be  set by the dissipation
coefficient $b$, but this estimate has not been tested experimentally.

The third of the predictions corresponds to observations in PMMA,
which observe the emergence of microcracks after about 40\% of the
Rayleigh wave speed. The first publication we are aware of describing
these as an important factor in the fracture of PMMA is by
Doyle \rr{Doyle}
{1983}; their role has also been emphasized by Ravi-Chandar
and Knauss \rr{Ravi}
{1984}.  Our use of thin strip samples enables us to obtain cracks
traveling in steady state at different mean velocities, and locate the
onset of micro-cracking with greater precision. A drawing of these
cracks soon after onset is displayed in \Fig{microscope}. We can use
\Eq{L93.11} to estimate the spacing of these branches.

The main experimental ingredient needed for the estimate is acoustic
attenuation at high frequencies, which has been measured in PMMA by Jackson,
Pentecost, and Powles \rr{Jackson_1972}
{1972}.
They find that attenuation per cycle is 
nearly constant in PMMA as a function of frequency and temperature,
and obeys
$$\alpha\lambda= 0.02-0.07,\EQN EO1$$
where $\alpha$ is attenuation per length, and $\lambda$ is the
wavelength of the waves one wants to consider (the experimental
measurements are for longitudinal waves.)
 The lower value of $0.02$
holds up to 100 MHz; there is a gap in the measurements, and $0.07$ is
measured at 10 GHz. Taking  $\alpha\lambda=0.02$, one has that
attenuation per time $b$ is given by $b\lambda/c_l=0.02$, with $c_l$
the longitudinal sound speed. Polymer units of size 0.5 $\mu$
separate coherently in PMMA  so PMMA
will be treated as a lattice of masses separated by $a=$0.5$\mu$m.
Now the waves
being excited by the passage of the crack have wavelength
$\lambda=ac_l/v$, where $a=$0.5$\mu$m is the lattice spacing and $v$ is
the crack velocity, so
$ b\sim 0.02 v/a \sim \rm 12\ MHz, $
for $a\sim 0.5 \mu$m  and $v\sim 300$ m/sec.
Taking $[\partial u_h/\partial\Delta] (\Delta-\Delta_c)/\delta u_h$ to
be on the order of 10\%, one has from 
\Eq{L93.11} and \Eq{EO1} the crude estimate that microcracks appear
with a frequency of  
$$6 {\rm MHz} \Rightarrow\  {\rm Spacing=}0.05{mm} \EQN EO2$$ 
for a crack traveling at 300 m/sec. 
Close examination of the 1/32 inch samples of PMMA reveals a comb of
micro-cracks, shown in \Fig{microscope}, resembling those in
\Fig{xiang}, and extending into the surface with a typical separation
of about 0.07 mm. This spacing is larger than the rough
theoretical estimate, but of the same order of magnitude.  Both in the
simulations and in the experiment, microcracks on the upper and lower
portions of the surface are staggered. 

There is also experimental evidence for the claim that the onset of
the microcracking instability is sensitive to details of lattice
structure. As shown in \Tbl{lattice.results}, moving from a triangular
to a square lattice pushes onset of the instability to higher
velocities; the general lesson should be that cleavage along weak
lattice planes discourages micro-cracking. Indeed, cracks reaching
the transverse wave speed (and in one case claimed to exceed it) have
been measured in cleavage of LiFl \rr{Gilman}
{Gilman, Knudsen, and Walsh, 1958}, while
Cotterell \rr{Cotterell_64}{1964} has measured  speeds approaching
the Rayleigh wave speed for cracks traveling along prescribed  grooves
in  PMMA.

The comparison between
simulation and experiment has to be made
with two qualifications. First, the simulations are on strips only
nine atoms high, from all points of view much smaller than the
experiments. Second, the experiments are clearly not two-dimensional. 
The cracks seen in \Fig{microscope} are those that were visible within
a particular plane, about 100 $\mu$ thick. Changing the vertical plane
changes the details of the pattern of microcracks; this observation
partly explains why the microcracks in the figure appear intermittent.

\figure{microscope}
\infiglist{mic}
\Caption Drawing of small microcracks that begin to
emerge from a crack after it passes a velocity of around 330 m/sec
in Plexiglas (PMMA). The basic geometry is the same as that shown in
\Fig{xiang}, with a spacing on the order of 0.07 mm. This diagram shows
the microcracks near the point where they first appear; later, they
are longer and more densely packed. The microcracks shown in the
figure are those that appear in a microscope whose depth of field is
about 100 $\mu$, and the pattern of microcracks changes as one looks
at different vertical planes by changing the focus of the microscope. 
\endCaption
\epsfysize=2truein
\centerline{\epsffile{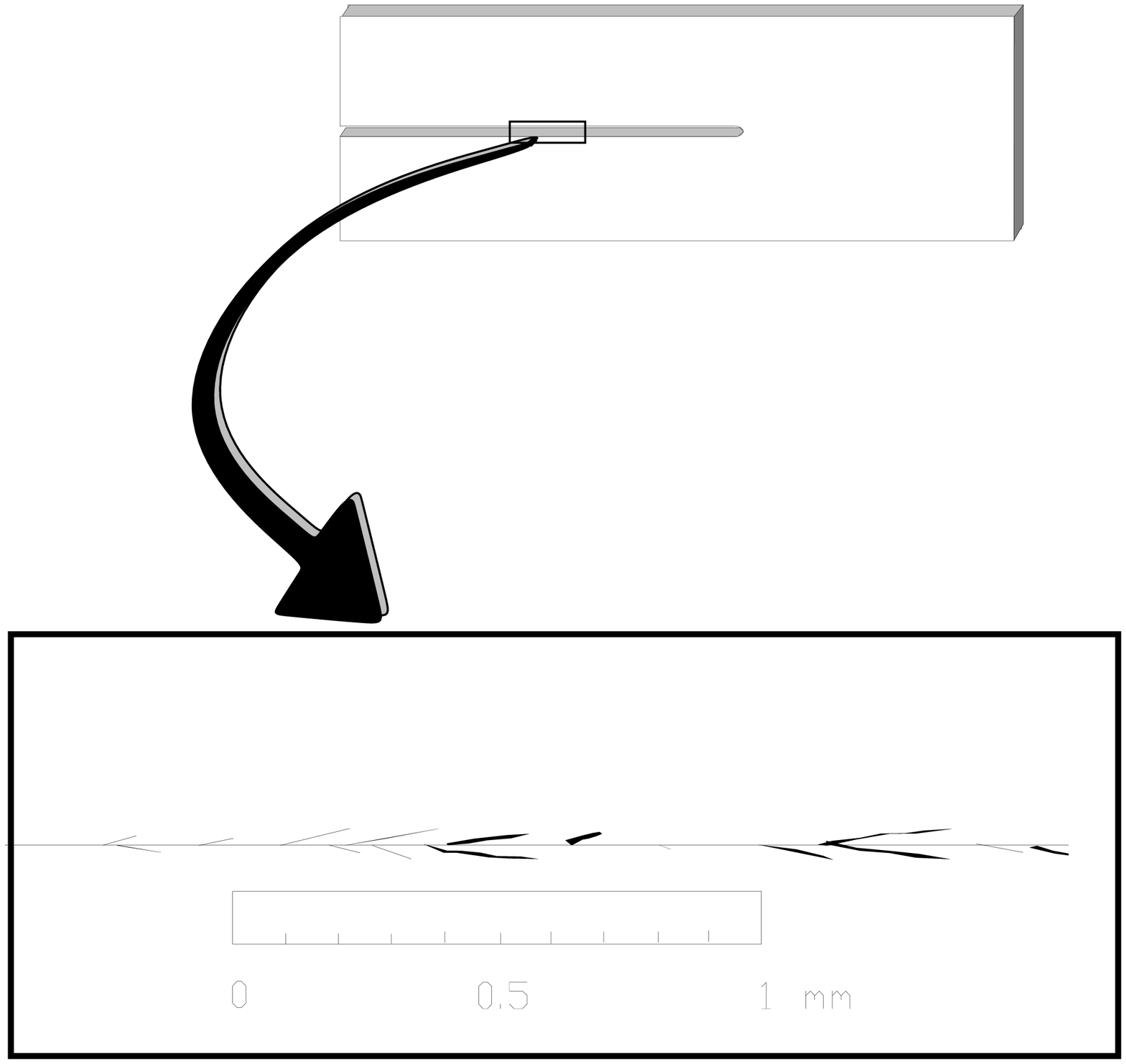}}
\endfigure

In connection with this three-dimensional structure, we note that
Perrin and Rice \rr{Perrin_94}
{1994} have recently found an interesting
implication of elastic theory. When a crack front is considered in
three dimensions, it is linearly stable against local fluctuations in
fracture toughness, but upon repeated contact with inhomogeneities,
the crack front can deform considerably, and experience velocity
fluctuations on the order of a quarter of the Rayleigh wave speed.
Therefore, a crack moving nominally at one-third the wave speed may
locally be moving much faster (or much slower). This picture provides
one possible explanation for the fact that microcracks appear
experimentally at velocities noticeably lower than that predicted by
the lattice models. 

In prior publications
\rr{Fineberg_91}
{Fineberg {\lowercase{\it et.~al.}}, 1991, 1992}
, our group has described a 500 KHz frequency emitted from PMMA,
corresponding to a 1 mm scale structure on 
the surface. This spatial scale is several times larger than the one
we have attributed to the onset of microcracking; the 1 mm
oscillations begin in earnest at a somewhat higher 
velocity (400 m/sec) than the initial microcracks (330 m/sec), and
correspond to bunches of especially dense microcracks grouped at roughly
1 mm intervals.  We do not yet know whether this phenomenon is the
natural evolution of the microcracking instability when driven to
large amplitudes, or whether new physical processes are coming into
play.  Thus the original oscillatory instability with which we began
our investigations is still not well understood.

While resolving numerous questions that arise in continuum models, and
providing a qualitative understanding for many features of  the
experiments, the lattice models raise as many questions as they answer.
When should one get micro-cracks, and when dislocations?  How do
dynamic ductile-brittle transitions occur? What are the effects of 
thermal noise? How is quasi-static motion possible?
What would happen in a random lattice? How should one treat a polymer?
How can the results be generalized to three dimensions? How does the
instability progress towards macroscopic branching? What happens in
larger-scale simulations with realistic bond forces? What are the
implications for fracture testing? These are some of the points that 
deserver further investigation.

This work was supported in part by the Texas Advanced Research Program, grant
number 3658--002. These topics were discussed with many of the
participants in an UCSB-ITP workshop on Spatially Extended
Nonequilibrium Systems, particularly J.~S.~Langer, and M.~Barber,
and are the result of a collaboration with experimentalists at the
Center for Nonlinear Dynamics at UT Austin, including
H.~Swinney,  J.~Fineberg, J.~Urbach, and W.~McCormick. Special thanks
are due to J.~Urbach for assistance with electronics in velocity
measurements. 

\nosechead{References}
{\sc{Ashurst, W.~T.}}, and {\sc{Hoover, W.~G.}}(1976) Microscopic
fracture studies in the two-dimensional triangular lattice,
 {\sl  Phys.~Rev.}   {\bf B14}, 1465-1473.

{\sc{Atkinson, W.}}, and {\sc{Cabrera, N.}}(1965) Motion of a
Frenkel-Kontorova dislocation in a one-dimensional crystal, {\sl
Phys.~Rev.}  {\bf 138}, A763-A766

{\sc{Bergkvist, H.~}}(1973) The motion of a brittle crack, {\sl
J.~Mech.~Phys.~Solids}    {\bf 21}, 229-239.

{\sc{Bergkvist, H.~}}(1974) Some experiments on crack motion and
arrest in polymethylmethacrylate,, {\sl  Engng.~Fracture~Mech.}   {\bf
6}, 621-626.

{\sc{Celli, V.}}, and {\sc{Flytzanis, N.}}(1970) Motion of a screw
dislocation in a crystal, {\sl  J.~Appl.~Phys.} {\bf 41}, 4443-4447.

{\sc{ Cotterell, B.}} (1965) Velocity effects in fracture propagation,
{\sl Appl.~Mater.~Res.} {\bf 4} 227-232.

{\sc{D\"oll, W.}}, and {\sc{Weidmann, G.~W.}}(1976) Transition from
slow to fast crack propagation in PMMA, {\sl J.~Mater.~Sci.~Lett.}   {\bf 11},
2348-2350.

{\sc{Doyle, M.~}}(1983)  A mechanism of crack
branching in polymethylmethacrylate and the origin of the bands on the
surfaces of fracture, {\sl  J.~Mater.~Sci.} {\bf 18}, 687-702.

{\sc{Fineberg, J., Gross, S., Marder, M.}} and {\sc{Swinney, H.}}, {\sc{{\it \lowercase{et.~al.}}}}(1991)
Instability in dynamic fracture, {\sl
Phys.~Rev.~Lett.}  {\bf 67}, 141-144.

{\sc{Fineberg, J., Gross, S., Marder, M.}} and {\sc{Swinney, H.}}, {\sc{{\it \lowercase{et.~al.}}}} (1992)
Instability in the propagation of fast cracks, {\sl
Phys.~Rev.}   {\bf B45}, 5146-5154.

{\sc{Freund, L.~B.}}(1974)Crack
propagation in an elastic solid subjected to general loading. IV.
Obliquely incident stress pulse,  {\sl  J.~Mech.~Phys.~Solids}  {\bf
22}, 137-146. 

{\sc{Freund, L.~B.}} (1990){\it Dynamic Fracture Mechanics}
Cambridge University Press, New York.

{\sc{Fuller, K.~N.~G.}}, {\sc{Fox, P.~G.~}}, and {\sc{Field, J.~E.~}}
(1975) The temperature rise at the tip of fast-moving cracks in glassy
polymers, {\sl Proc.~R.~Soc.~Lond.~A}   {\bf 341}, 537-557.

{\sc{Gao, H.~}}(1993) Surface roughening and branching instabilities
in dynamic fracture, {\sl  J.~Mech.~Phys.~Solids}   {\bf 41}, 457-486.

{\sc{Gilman, J.~J.}}, {\sc{Knudsen, C.}}, and {\sc{Walsh, W.~P.~}}
(1958) Cleavage cracks and dislocation in LiF crystals, {\sl
J.~Appl.~Phys}   {\bf 29}, 601-607.

{\sc{Gradshteyn I.~S.}} and {\sc{Ryzhik, I.~M.}} (1980) Table of
integrals, series and products. Academic Press, New York.

{\sc{Green, A.~K.}}, and {\sc{Pratt, P.~L.}}(1974) Measurement of the
dynamic fracture toughness of polymethylmethacrylate by high-speed
photography, {\sl   Engng.~Fracture Mech.}  {\bf 6}, 71-80.

{\sc{Gross, S., Fineberg, J., McCormick, W.~D., Marder, M.,}} and
{\sc{Swinney, H.}} (1993) Acoustic
emissions from rapidly moving cracks, {\sl 
Phys.~Rev.~Lett.}  {\bf 71}, 3162-3165.

{\sc{Jackson, D.~A.}}, {\sc{Pentecost, H.~T.~A.}}, and {\sc{Powles,
J.~G.~}} (1972) Hypersonic absorption in amorphous polymers by light
scattering,  {\sl Mol.~Phys.}   {\bf 23}, 425-432.

{\sc{Kanninen, M.~F.}}, and {\sc{Popelar, C.}}(1985) {\it Advanced 
Fracture Mechanics}  Oxford University Press, New York.

{\sc{Katsamanis, F.~G.}}, and {\sc{Delides, C.~G.}}(1988) Fracture
surface energy measurements of PMMA: a new experimental approach,  {\sl
J.~Phys.}   {\bf D 21}, 79-86.

{\sc{Knauss, W.~G.}}(1966) Stresses in an infinite strip containing a
semi-infinite crack, {\sl  J.~Appl.~Mech.} {\bf 33}, 356-362.

{\sc{Kobayashi, A.}}, {\sc{Ohtani, N.}}, and {\sc{Sato, T.~}} (1974)
Phenomenological aspects of viscoelastic crack propagation, {\sl
J.~Appl.~Polymer Sci.}   {\bf 18}, 1625-1638.

{\sc{Kulakhmetova, Sh.~A.}},  {\sc{Saraikin, V.~A.}}, and
{\sc{Slepyan, L.~I.}} (1984) Plane problem of a crack in a lattice,
{\sl Mechanics of Solids}   {\bf 19}, 101-108.

{\sc{Langer, J.~S.~}}(1993) Dynamic model of onset and propagation of
fracture, {\sl Phys.~Rev.~Lett.}   {\bf 70}, 3592-3594.

{\sc{Langer, J.~S.}}, and {\sc{Nakanishi, H.~}}(1993) Models of crack
propagation. II. Two-dimensional model with dissipation on the
fracture surface, {\sl Phys.~Rev.}   {\bf E48}, 439-448.

{\sc{Liu,  X.}} (1993) Dynamics of fracture propagation,  (Dissertation,
University of Texas)

{\sc{Liu, X.}}, and {\sc{Marder, M.}}(1991) The energy of a
steady-state crack in a strip, {\sl  J.~Mech.~Phys.~Sol}
{\bf 39}, 947-961.

{\sc{Machov\'a, A.}} (1992) Molecular dynamic simulation of microcrack
initiation,  {\sl Materials Science and Engineering}
{\bf A149} 153-165.

{\sc{Manneville, P.}} (1990) {\it Dissipative Structures and Weak Turbulence}, 
Academic Press, Boston Chapter 6, section 6

{\sc{Marder, M.~}}(1991) New dynamical equation for cracks, {\sl
Phys.~Rev.~Lett.}   {\bf 66}, 2484-2487.

{\sc{Marder, M.}}, and {\sc{Liu, X.~}}(1993) Instability in lattice
fracture, {\sl  Phys.~Rev.~Lett.} {\bf 71}, 2417-2420.

{\sc{Mecholsky, J.~J.}} (1985) Fracture analysis of glass surfaces,
in {\it Strength of Inorganic Glass}, ({\sc{C.~R.Kurkjian}}, ed.),
pp.~569-590. Plenum Press, New York.

{\sc{Noble, B.}}  (1959) {\it Methods Based on the 
Wiener-Hopf Technique  for the Solution of Partial 
Differential Equations} Pergamon Press, New York.

{\sc{Perrin, G.}},  and {\sc{Rice, J.~R.}} (1994), Disordering of a
dynamic planar crack front in a model elastic medium of randomly
variable toughness, {\sl J.~Mech.~Phys.~Solids} {\bf 42} 1047-1064.

{\sc{Ravi-Chandar, K.~}}, and {\sc{Knauss, W.~G.~}}(1984) An
experimental investigation into dynamic fracture: III. On steady state
crack propagation and crack branching, {\sl
Int.~J.~Fracture}   {\bf  26 }, 141-154.

{\sc{Rice, J.~R., Ben-Zion, Y.}},  and {\sc{Kim, K}}. (1994) Three-dimensional
perturbative solution for a dynamic planar crack moving unsteadily in
a model elastic solid,  {\sl J.~Mech.~Phys.~Solids} {\bf 42}, 813-843.

{\sc{Sieradzki, K., Dienes, G.~J., Paskin, A.,}} and
{\sc{Massoumzadeh, B.}}, (1989) 
Atomistics of crack branching, {\sl Acta.~Metall.} {\bf 36}, 651

{\sc{Slepyan, L.~I.}} (1981) Dynamics of brittle fracture in lattice,
{Doklady Soviet Phys.} {\bf 26} 538-540.

{\sc{Slepyan, L.~I.}}(1992) Principle of maximum energy dissipation rate
in crack dynamics, {\sl  Doklady Soviet Phys.}   {\bf 41}, 1019-1033.

{\sc{Slepyan, L.~I.}}(1993) Principle of maximum energy dissipation rate
in crack dynamics, {\sl  J.~Mech.~Phys.~Solids}   {\bf 41}, 1019-1033.

{\sc{Takahashi, K.}}, {\sc{Matsushige, K.~}} and {\sc{Sakurada, Y.~}},
(1984) Precise evaluation of fast fracture velocities in acrylic
polymers at the slow-to-fast transition, {\sl  J.~Mater.~Sci.}  {\bf
19}, 4026-4034.

{\sc{Thomson, R.~}}(1986) Physics of fracture, {\sl  Solid State
Physics}   {\bf 39}, 1-129.

{\sc{Thomson, R.}}, {\sc{Hsieh, C.}}, and {\sc{Rana, V.}}(1971)
Lattice trapping of fracture cracks, {\sl J.~Appl.~Phys.} {\bf 42},
3154-3160. 

{\sc{Washabaugh, P.~D.}}, and {\sc{Knauss, W.~G.}}(1993) Non-steady
periodic behavior in the dynamic fracture of PMMA,  {\sl
Int.~J.~Fracture}   {\bf 59}, 189-197.

{\sc{Willis, J.~R. }}  (1990) Accelerating cracks and related
problems, , in {\it Elasticity: Mathematical Methods
and Applications} ({\sc{G.~Eason, and R.~W.~Ogden}}, eds.)
pp.~397-409. Halston Press, New York.

{\sc{Xu, Y.}}, and {\sc{Keer, L.~M.}}(1992) Non-planar deviation of an
initially straight moving crack,  {\sl Engng.~Fracture~Mech.} {\bf
41}, 577-585. 

{\sc{Yoffe, E.~H.}}(1951) The moving Griffith crack, {\sl  Phil.~Mag.}
{\bf  42}, 739-750.

{\sc{Zhou, S.~J., Carlsson, A.~E.}}, and  {\sc{Thomson, R.}}(1994) Crack
blunting effects on dislocation emission from cracks,  {\sl
Phys.~Rev.~Lett.}   {\bf 72}, 852-855.

\Appendix{I} {Calculation of Lattice-Trapped States}

In this Appendix, we will calculate the static solutions of \Eq{OD1}.
Dropping the subscript $+$, one begins with the equation
$$0=u_{m+1}-2u_m+u_{m-1}+{1\over N}(U_N-u_m)-2u_m\theta_m,\EQN
LT1$$
where
$$\theta_m=\cases{1& for $m\geq 0$\cr 0 & for $m<0$}.\EQN LT2$$
For this solution to follow from \Eq{OD1}, one must have that
$$u_0<1{\rm\ and\ } u_{-1}>1.\EQN LT3$$
In other words, the bond at zero has not snapped, so it must be stretched
less than a distance of $1$, but the bond at $-1$ has  snapped, so
it must be stretched more than $1$.
The solution of \Eq{LT1} is
$$u_m=\cases{ u_r(m)={U_N\over 2N+1}+u_r y_r^m& for $m\geq 0$\cr
              u_l(m)=U_N+ u_l y_l^m& for $m<0$},\EQN LT3$$
where
$$y_l-2+1/y_l-1/N=0,\EQN LT4;a$$
and
$$y_r-2+1/y_r-2-1/N=0.\EQN LT4;b$$
In order for the left and right solutions to match onto one another smoothly,
one must demand that
$$u_l(0)=u_r(0) {\rm \ and \ } u_l(-1)=u_r(-1).\EQN LT5$$
The four equations \Eq{LT4} and \Eq{LT5} determine $u_l$, $u_r$, $y_l$
and $y_r$ as  functions of $U_N$. Using $U_c=\sqrt{2N+1}$, 
turning the condition \Eq{LT2} into a condition on $\Delta$, and
working in the limit $N\rightarrow\infty$ gives
$$\Delta<{\sqrt 3+1\over\sqrt 2}\ {\rm  for\ } u_0<1,\EQN LT6;a$$
and
$$\Delta>{\sqrt 3-1\over\sqrt 2}\ {\rm  for\ } u_{-1}>1\EQN LT6;b$$
These are the boundaries for the region of lattice trapping.

\Appendix{II}{Evaluation of $\Delta$ for small and for large $v$}
In this Appendix, we will evaluate $\Delta(v)$ for the one-dimensional
model in the limits $v\rightarrow 0$ and $v\rightarrow 1$, starting
from \Eq{OD20}

It is possible to evaluate the product
$$P(A)=\prod{w_i^+ \over w_i^- },\EQN PF7$$ 
where for generality $w_i$ is the root of
$$\omega^2-4\sin^2(\omega/2v)-4A^2\EQN PF8$$
analytically in certain limits. In the limit of low velocity, it is
helpful first to write the condition that \Eq{PF8} have a root as
$$\omega=2v\sin^{-1}\sqrt{\omega^2/4-A^2}.\EQN PF9$$
In the limit of low velocity, by looking at a graph of \Eq{PF8}, one
sees that the roots are given approximately by
$$ w_i^+=2v[(n_0+i)\pi -\sin^{-1}(\sqrt{v^2[n_0+i]^2\pi^2-A^2})],\EQN PF10;a$$
$$ w_i^-=2v[(n_0+i)\pi +\sin^{-1}(\sqrt{v^2[n_0+i]^2\pi^2-A^2})],\EQN
PF10;b$$
and that there is always one more positive root $w_i^+$ than there are
positive roots $w_i^-$. The starting integer $n_0$ is given by 
$$A=2vn_0\pi\EQN PF11;a$$
(one will have to fiddle around with $v$ a bit to make this precisely
true) and the largest value of $i$ is $n_f$, given by
$$\sqrt{A^2+1} =(n_0+n_f)v\pi\EQN PF11;b$$
(this comes close to the truth in the limit of small $v$). 
 
For the moment, let us restrict ourselves just to the positive roots
of \Eq{PF8}. Then
$${1\over 2} \ln P(A)=\sum_{i=0}^{n_f-1}
\ln[{(n_0+i)\pi-\sin^{-1}(\sqrt{v^2[n_0+i]^2\pi^2-A^2}) \over
(n_0+i)\pi+\sin^{-1}(\sqrt{v^2[n_0+i]^2\pi^2-A^2})
}]+\ln[(n_0+n_f)\pi] \EQN PF12$$
$$=\sum_{i=0}^{n_f-1}
\ln[{1-\sin^{-1}(\sqrt{v^2[n_0+i]^2\pi^2-A^2})/(n_0+i)\pi \over
1+\sin^{-1}(\sqrt{v^2[n_0+i]^2\pi^2-A^2})/(n_0+i)\pi
}]+\ln[(n_0+n_f)\pi] \EQN PF13$$
$$=\sum_{i=0}^{n_f-1}
(-2)\sin^{-1}(\sqrt{v^2[n_0+i]^2\pi^2-A^2})/(n_0+i)\pi
+\ln[(n_0+n_f)\pi] \EQN PF14$$
$$=\int_A^{\sqrt {A^2+1}}dx ({-2\over \pi}) {\sin^{-1}\sqrt{x^2-A^2}\over x}
+ \ln(\sqrt{A^2+1}/v)\EQN PF15$$
and integrating by parts gives
$${1\over 2}\ln P(A)= \ln{1\over v} +{2\over \pi} \int_A^{\sqrt{A^2+1}}
{x\ln x dx\over \sqrt{x^2-A^2}\sqrt{1+A^2-x^2}}, \EQN PF16$$
which after the change of variables $x^2=A^2+\sin^2\theta$ becomes
$$P(A)={1\over v^2} \exp[{1\over\pi}\int_0^\pi \ln[A^2+\sin^2\theta]]d\theta.
\EQN PF17$$
In the particular case of \Eq{OD20}, one wants to evaluate 
$$({\Delta})^2={P(\sqrt{1+\epsilon^2}/2)\over
P(\epsilon/2)}
=\exp[{1\over\pi}\int_0^\pi \ln[{\epsilon^2/4+1/4+\sin^2\theta \over
\epsilon^2/4+\sin^2\theta }]]d\theta.
\EQN PF18$$
Gradshteyn and Rizhyk (1980) have the integral (4.399) 
$$\int_0^{\pi} dx \ln(1+a\sin^2x)=2\pi\ln({1+\sqrt{1+a}\over 2}).\EQN
PF18.5$$
This gives finally
$${\Delta
}={\sqrt{\epsilon^2/4+1/4}+\sqrt{1+\epsilon^2/4+1/4} \over
\epsilon/2+\sqrt{1+\epsilon^2/4} }.\EQN PF18.6$$
In the limit $\epsilon\rightarrow 0$, one has that
$${\Delta}={1+\sqrt 5\over 2}=1.6180\dots,\EQN PF18.7$$
the golden mean, in agreement to three places with the direct
evaluation of the roots in \Eq{OD20}.
One has that a stationary lattice crack in a noiseless environment
will not begin to move until the driving strain exceeds by this amount
the strain that would be predicted in a continuum model.

At velocities that approach 1, there is one root $f^+$ and one root
$g^+$. In the limit of small $\epsilon$, and for $v=1$, $g^+=1.91892$.
One finds $f^+=\sqrt{2\sqrt{3}\epsilon+12(1-v)}$.
Therefore
$$ \Delta\rightarrow \sqrt{1.91892\over
\sqrt{2\sqrt{3}\epsilon+12(1-v)}}.\EQN PF18$$

\Appendix{III}{Calculation of $\Delta$ for small damping}
This Appendix describes the calculation of $\Delta(v)$ in the limit of
zero damping, $b=0$ for the triangular lattice of Section 3. Begin
with \Eq{OD20}. 

In order to make use of this expression,  one must find an efficient
way to locate 
the roots and poles of $Q$. This task is equivalent to finding the
roots of $F$, defined in \Eq{L93.5.1b}, and the roots of
$$G(\omega)\equiv F(\omega)-1-\cos (\omega/2v),\EQN AP1.4$$
which is the denominator of $Q$. Both $F$ and $G$ have poles, but at
the same values of $\omega$, and these poles cancel. The roots fall
into two classes. In the first class are roots for which $|z|>1$.
These roots are well separated, and can be located by a standard
root-finding algorithm. In the second class are roots for which
$|z|\leq 1$. These roots are very finely spaced, and for each region
where $|z|<1$, there are of order $N$ roots. These may be located by
looking for the values of $y$
where $F$ vanishes. These are denoted by $y^F$, and are indexed by $j$,
which varies from $1$ to $N$. Inverting $F$ to find $y$ one has
$$y_{j}^F=e^{i(2j-1)\pi/(2N+1)+\ln[{y_{j}^F-\cos(\omega/2v)\over
y_{j}^F\cos (\omega/2v)-1}]/(2N+1)}.\EQN TL 17$$
Therefore
$$z_{j}^F=\cos[{(2j-1)\pi\over 2N+1}+{1\over
i(2N+1)}\ln\lb{y^F_{j}-\cos \omega/2v\over
y^F_{j}(\cos\omega/2v)-1}\rb] \EQN TL18;a$$
Similarly, inverting $G$ to find $y$ and $z$ one gets
$$z_{j}^G=\cos[{2j\pi\over 2N+1}],\EQN TL18;b$$
The roots of $G$ are therefore very easy to find. By finding solutions
of \Eq{TL18;a} one finds the roots of $F$. Once all the roots are
located, one determines the sign of their imaginary part for
infinitesimal $b$, and finally inserts them into \Eq{OD20} to find
$\Delta$. 
\Appendix{IV}{Linear stability of steady state solutions}
This Appendix carries out the linear stability analysis of steady
state solutions for the Mode III model in detail.
To begin, it is useful to recall the symmetries of the steady
state solutions. These are
$$u^0(m,n,t)=u^0(m+1,n,t+1/v)\EQN TLP3;a$$
and also
$$u^0(m,n,t)=-u^0(m,-n,t-[1/2-g_n]/v).\EQN TLP3;b$$
Therefore the perturbations can be taken first to be eigenfunctions
of the operator which translates $m$ by 1, and $t$ simultaneously by
$1/v$, perturbations with time dependence of
the form $e^{qt}f(n,t-m/v)$. Second, the perturbations are eigenfunctions of
the operator which inverts states around the $x$ axis while moving $t$
forward by $1/2v$.  If this operation is repeated twice, it simply
produces one of the translations of the first operator, and must
therefore have eigenvalue $e^{qt}$. The operation carried out only
once must therefore have eigenvalue $\pm e^{qt/2}$; even and odd
modes, associated with the inversion symmetry.

Therefore the states will be of the form
$$u^0(n,t-m/v)+u^1(n,t-m/v)e^{qt},\EQN LP1$$
with $u^1$ small, and with both even and odd inversion symmetries
possible. 

Placing this state into \Eq{L93.1;a} and expanding to first order, one has
that 
$$\eqalign { &{q^2 u^1(m,n)+2q\dot u^1(m,n)+\ddot u^1(m,n)}=\cr
&{1\over 2}[\eqalign{+u^1(m+g_{n+1}-1,n+1)&+u^1(m+g_{n+1},n+1)\cr
+u^1(m-1,n)-6&u^1(m,n)+u^1(m+1,n)\cr  
 +u^1(m+g_{n-1}-1,n-1)+&u^1(m+g_{n-1},n-1)}]\cr -&b{\dot u^1}(m,n)}\EQN LP2;a$$
if $n>1/2$, taking $u_f=1$
$$\eqalign
 { &{q^2 u^1(m,n)+2q\dot u^1(m,n)+\ddot u^1(m,n)}=\cr
&{1\over 2}[\eqalign{+u^1(m+g_{n+1}-1,n+1)&+u^1(m+g_{n+1},n+1)\cr
+u^1(m-1,n)-4&u^1(m,n)+u^1(m+1,n)\cr  
 +[u^1(m,n-1)-u^1(m,n)]&[\theta(-t)-\delta(1-u^0(t))]\cr
+[u^1(m+1,n-1)-u^1(m,n)]&[\theta(1/2v-t)-\delta(1-u^0(t-1/2v))]} 
]\cr
&-b{\dot u^1}(m,n)}\EQN LP2;b$$ 
if $n=1/2$. The boundary condition is that $u^1$ vanish at
$n=\pm(N+1/2)$, and $u^0$ is defined by \Eq{TL12} as the difference
between two mass points above and below the crack line in the steady
state. Upon Fourier transforming \Eq{LP2} one sees that it differs
from the solution of \Eq{L93.1} only in two ways. First, $-\omega^2\rightarrow
(q-i\omega)^2$, changing the definition of $z$ in \Eq{TL8}. Second,
the driving force on the crack face is no longer the applied stress
$\sigma_\infty$, but is now a set of two delta functions appearing
near the end of \Eq{LP2;b}.

Fourier transforming \Eq{LP2} one has therefore that
$$z_q={3-\cos(\omega/v)+(q-i\omega)^2-ib\omega\over 2\cos(\omega/2v)}\EQN
LP3;a$$ 
$$y_q=z_q+\sqrt{z_q^2-1},\EQN LP3;b$$
and
$$\eqalign{&u^1_{1/2}(\omega)F_q(\omega)\cr
&-{1\over 2}\int dt e^{i\omega t}
 \lb[u^1_{-1/2}(t)-u^1_{1/2}(t)]\theta(-t)+[u^1_{-1/2}(t-1/v)-
u^1_{1/2}(t)]\theta(1/2v-t)\rb \cr  
= -& {[u^1_{-1/2}(0)-u^1_{1/2}(0)]+e^{i\omega/2v} [u^1_{-1/2}(-1/2v)-u^1_{1/2}(1/2v)]\over 2 \dot
u^0(0) }}\EQN LP4$$
with
$$F_q(\omega)=\lb{ y_q^{[N-1]}-y_q^{-[N-1]}\over
y_q^{N}-y_q^{-N}}-2z_q\rb \cos(\omega/2v)+1\EQN LP4.1$$
At this point, one needs to distinguish between the modes which are
odd and those which are even under inversion. Define
$$U(t)={u^1_{1/2}(t)-u^1_{-1/2}(t)\over 2}\EQN LP5;a$$
for the modes such that 
$$u^1_{1/2}(t)=-u^1_{-1/2}(t-1/2v),\EQN LP5;b$$
and
$$V(t)={u^1_{1/2}(t)-u^1_{-1/2}(t)\over 2}\EQN LP6;a$$
for the modes such that
$$u^1_{1/2}(t)=u^1_{-1/2}(t-1/2v),\EQN LP6;b$$
The least stable mode is of the type  $U(t)$, which will therefore be
of interest in what follows.

One has that
$$U(\omega)F_q-(1+\cos\omega/2v)U^-=-(1+\cos\omega/2v) {U(t=0)\over \dot
u^0(0)} \EQN LP7$$
or
$$U^+Q_q+U^-=(1-Q_q){U(t=0)\over \dot u^0(0)}\EQN LP8$$
with $Q_q$  given by \Eq{L93.5.1}; the subscripts $q$ are meant
to remind that $z$ depends upon $q$. This may be rewritten
$$U^+{1\over Q_q^+}+U^-{1\over Q_q^-}
=({1\over Q_q^-}-{1\over Q_q^+}){U(t=0)\over
\dot u^0(0)} .\EQN LP9$$
This expression can be decomposed immediately to give
$$U^-(\omega){1\over Q_q^-(\omega)}=({1\over Q_q^-}-{1\over
Q_q^-(\infty)} ){U(t=0)\over \dot u^0(0)} ,\EQN LP10$$
where a constant term has been included to avoid a delta function
singularity in $U(t)$ at $t=0$. Since as before  $Q_q^-(\omega)$ goes to one at
when $\omega$ goes to infinity
$$U^-(\omega)=(1-{Q_q^-}){U(t=0)\over
\dot u^0(0)} .\EQN LP10$$

Because of time translation invariance, one expects there to be a zero
mode, whose eigenfunction is $\dot u^0(t)$. By comparison with
\Eq{L93.6}, one sees immediately that this is the case, since 
$Q_{q=0}$ appearing in  \Eq{LP10}
is identical to the function $Q$ defined previously by \Eq{L93.5.1}.
The general consistency condition which tells
which values of $q$ are allowed is
$$U^-(t=0)=\lim_{\epsilon\rightarrow 0} \int {d\omega\over 2\pi}
e^{i\omega\epsilon} (1-{Q_q^-(\omega)}){U(t=0)\over \dot u^0(0)} \EQN LP10$$
$$\Rightarrow  \dot u^0(0)=\int {d\omega\over 2\pi} 
(1-{Q_q^-(\omega)}). \EQN LP11$$
Using the fact that that the zero mode is just $\dot u^0$, 
one can also write
$$0=\int {d\omega\over 2\pi} 
({Q_0^-(\omega)}-{Q_q^-(\omega)})\equiv S_q-S_0. \EQN LP12$$ 
When this equation has a solution for $q$ with positive real part the
corresponding steady state is unstable, and otherwise it is stable.

\Appendix{V}{Cracks which go faster when pulled harder are stable}

This Appendix derives relation \Eq{LP26}.

Notice from \Eq{LP3;a} that
$$ z_q(a\omega,av,b/a,q)=z_q(\omega,v,b,q-ia\omega+i\omega)\EQN LP14;a$$
Therefore
$$Q_q(a\omega,av,b/a,q)=Q_q(\omega,v,b,q-ia
\omega+i\omega)\EQN LP14;b$$  
(since the additional dependence of $Q_q$ upon $\omega$ and $v$ is of
the form $\omega/v$)
$$\Rightarrow \para Q_q,q,=i\para Q_q,\omega,-{v\over i\omega} \para Q_q,v,
+{b\over i\omega} \para Q_q,b,.\EQN LP15$$
Now
$$Q_q^-(\omega)=\exp[\int_{-\infty}^0 dt e^{i\omega t}\int
{d\omega^\prime \over 2\pi} e^{-i\omega^\prime t} \ln
Q_q(\omega^\prime)] ,\EQN LP17$$
so
$$\para Q^-_q,q,=Q^-_q \int_{-\infty}^0 dt e^{i\omega t}\int
{d\omega^\prime \over 2\pi} e^{-i\omega^\prime t} \para ,q, \ln
Q_q(\omega^\prime) ,\EQN LP18$$
which implies, since $Q_q$ depends upon $q$ only through $z$, that
$$\para
Q^-_q,q,=Q^-_q \int_{-\infty}^0 dt e^{i\omega t}\int
{d\omega^\prime \over 2\pi} e^{-i\omega^\prime t}  \lb i\para
,\omega^\prime ,-{v\over i\omega^\prime} \para ,v,
+{b\over i\omega\prime} \para ,b, \rb \ln
Q_q(\omega^\prime) .\EQN LP19$$

Considering the terms in \Eq{LP19} one by one, we have first
$$\int_{-\infty}^0 dt e^{i\omega t}\int
{d\omega^\prime \over 2\pi} e^{-i\omega^\prime t} i\para
,\omega^\prime, \ln Q_q\EQN LP20;a$$
$$=\int_{-\infty}^0 dt e^{i\omega t}\int
{d\omega^\prime \over 2\pi} e^{-i\omega^\prime t} i(it) \ln Q_q\EQN
LP20;b$$
$$=i\para ,\omega, \int_{-\infty}^0 dt e^{i\omega t}\int
{d\omega^\prime \over 2\pi} e^{-i\omega^\prime t}  \ln Q_q\EQN
LP20;c$$
$$=i\para ,\omega, \ln Q_q^-(\omega)\EQN
LP20;d$$

Second,
$$\int_{-\infty}^0 dt e^{i\omega t}\int
{d\omega^\prime \over 2\pi} e^{-i\omega^\prime t} {1\over i
\omega^\prime} \ln Q_q\EQN LP21;a$$
$$=\lim_{\epsilon\rightarrow 0} \int {d\omega\over 2\pi} {e^{-i\epsilon
\omega^\prime} \over i(\omega-\omega^\prime)}{1\over i\omega^\prime}
\ln Q_q^-\EQN LP21;b$$
$$=\lim_{\epsilon\rightarrow 0} \int {d\omega\over 2\pi} {e^{-i\epsilon
\omega^\prime} \over i\omega}[{1\over i(\omega-\omega^\prime)}+{1\over
i\omega^\prime}] 
\ln Q_q^-\EQN LP21;c$$
$$={1\over i\omega} \ln({Q^-_q(\omega)\over Q_q^-(0)})\EQN LP21;d$$
Putting together \Eq{LP20} and \Eq{LP21} gives that
$$\para ,q, \ln Q_q^-(\omega)=[i\para ,\omega,-{v\over i\omega} \para
,v, +{b\over i\omega} \para ,b,] \ln ({Q_q^-(\omega)\over
Q_q^-(0)}).\EQN LP22$$ 
Returning with this result to \Eq{L93a.7.1} gives 
$$S^\prime_0=\int {d\omega\over 2\pi} Q_0^-(\omega) {v\over
i\omega} \para ,v, \ln ({Q_0^-(\omega)\over
Q_0^-(0)}).\EQN LP23$$
Since
$$Q^-_q(\infty)=1\EQN LP24$$
one has finally that
$$S^\prime_0=-v\para ,v, \ln Q_0^-(0)=v\para ,v, \ln {u(0)\over
\Delta} \EQN LP25$$
$$\Rightarrow S^\prime_0=-\para \ln\Delta,\ln v, \EQN LP26$$

An additional expression makes it possible to evaluate $S_q$ more
quickly, and without needing recourse to Fourier transforms. One has
that
$$\int {d\omega\over 2\pi} [Q^-_q(\omega)-1]=\lim_{\omega\rightarrow
\infty} i\omega  [Q^-_q(\omega)-1].\EQN LP27$$
$$=\lim_{\omega\rightarrow \infty} i\omega \lb
\exp[\int_{-\infty}^0 dt e^{i\omega t}\int
{d\omega^\prime \over 2\pi} e^{-i\omega^\prime t} \ln
Q_q(\omega^\prime)]-1\rb .\EQN LP28$$

The large $\omega$ behavior in the exponential is given by the
discontinuity in the function of $t$ at $0$, so one has next
$$\lim_{\omega\rightarrow \infty} i\omega \lb
\exp[{1\over i\omega} \int
{d\omega^\prime \over 2\pi} \ln
Q_q(\omega^\prime)]-1\rb \EQN LP29$$
$$= \int{d\omega^\prime \over 2\pi} \ln
Q_q(\omega^\prime) .\EQN LP30$$
One has finally that
$$S_q= \int{d\omega^\prime \over 2\pi} \ln [{Q_0(\omega^\prime)\over
Q_q(\omega^\prime)}] .\EQN LP30$$

\Appendix{VI}{Calculation of $Q$ for fully two-dimensional model}

This Appendix describes the calculation of $Q$ for the model of
Section \use{sec.modeI}. There seems no simple way to report it, and
the calculations were all carried out in MAXIMA; the batch file is
available upon request from the first author.

Let $\Delta_i$ describe the displacements between points as shown in
\Fig{triang}. The restoring force parallel to the direction of
equilibrium bonds will be $\kp$, while that perpendicular to this
direction will be $\kr$. The force due to the displacement of the
particle along $\vec \Delta_1=\vec u_{i-1,j+1}-\vec u_{i,j}$ is
$$\kp \hat d_{\|1}(\vec \Delta_1\cdot\hat d_{\|1})+\kr
\hat d_{\bot 1}(\vec\Delta_1\cdot\hat d_{\bot 1}) \EQN TR0.1$$
$$=\kp  ({-1\over 2},{\sqrt 3\over 2})({-1\over
2}\Delta_1^x,{\sqrt 3\over 2} \Delta_1^y)+\kr
({\sqrt 3\over 2},{1\over 2})({\sqrt 3\over 2}\Delta_1^x,{1\over 2}
\Delta_1^y) \EQN TR0.2$$
Adding up contributions from other particles in this way we get for the
force due to neighbors
$$\vec F(m,n)=\sum_{j=1}^6\sum_{q=\|,\bot}k_q\hat d_{qi}(\vec
\Delta_i(m,n)\cdot\hat d_{qi})\EQN TR0.3$$

In steady state, for $n>1/2$ the forces become
$$F^x=\eqalign{ ({1\over 4}\kp+{3\over
4}\kr)&[\eqalign{&u^x_{{n+1}}(t-(g_{n+1}-1)a/v)+u^x_{n+1}(t-g_{n+1}a/v)\cr
+&u^x_{n-1}(t-(g_{n+1}-1)a/v) +  u^x_{n-1}(t-g_{n+1}a/v)} -4u^x_{n}(t)]\cr
+{\sqrt 3\over   4}(\kr-\kp) &[\eqalign{&u^y_{n+1}(t-(g_{n+1}-1)a/v)-u^y_{n+1}(t-g_{n+1}a/v)\cr -&u^y_{n-1}(t-(g_{n+1}-1)a/v) +
u^y_{n-1}(t-g_{n+1}a/v)}] \cr
+\kp&[u^x_{n}(t+a/v)+u^x_{n}(t-a/v)-2u^x_{n}(t)]}\EQN TR11;a$$
$$F^y=\eqalign{ ({1\over 4}\kr+{3\over
4}\kp)&[\eqalign{&u^y_{{n+1}}(t-(g_{n+1}-1)a/v)+u^y_{n+1}(t-g_{n+1}a/v)\cr
+&u^y_{n-1}(t-(g_{n+1}-1)a/v) +  u^y_{n-1}(t-g_{n+1}a/v)} -4u^y_{n}(t)]\cr
+{\sqrt 3\over   4}(\kr-\kp) &[\eqalign{&u^x_{n+1}(t-(g_{n+1}-1)a/v)-u^x_{n+1}(t-g_{n+1}a/v)\cr -&u^x_{n-1}(t-(g_{n+1}-1)a/v) +
u^x_{n-1}(t-g_{n+1}a/v)}] \cr
+\kr&[u^y_{n}(t+a/v)+u^y_{n}(t-a/v)-2u^y_{n}(t)]}\EQN TR11;b$$
In \Eq{TR11}, all the displacements $u$ are evaluated at $m=0$, and this
index is therefore dropped, so that the only remaining index refers to
the layer number $n$. 
For $n=1/2$, one has instead
$$F^x=\eqalign{ ({1\over 4}\kp+{3\over
4}\kr)&[\eqalign{&u^x_{{n+1}}(t-(g_{n+1}-1)a/v)+u^x_{n+1}(t-g_{n+1}a/v)\cr
+&\theta(-t)(u^x_{n-1}(t-(g_{n+1}-1)a/v)-u_n^x(t))\cr +
&\theta(a/2v-t)(u^x_{n-1}(t-g_{n+1}a/v)-u^x_n(t))} -2u^x_{n}(t)]\cr
+{\sqrt 3\over   4}(\kr-\kp) &[\eqalign{&u^y_{n+1}(t-(g_{n+1}-1)a/v)-u^y_{n+1}(t-g_{n+1}a/v)\cr -&\theta(-t)(u^y_{n-1}(t-(g_{n+1}-1)a/v)-u^y_n(t))\cr +
&\theta(a/2v-t)(u^y_{n-1}(t-g_{n+1}a/v)-u^y_n(t))}] \cr
+\kp&[u^x_{n}(t+a/v)+u^x_{n}(t-a/v)-2u^x_{n}(t)]}\EQN TR12;a$$
$$F^y=\eqalign{ ({1\over 4}\kr+{3\over
4}\kp)&[\eqalign{&u^y_{{n+1}}(t-(g_{n+1}-1)a/v)+u^y_{n+1}(t-g_{n+1}a/v)\cr
+&\theta(-t)(u^y_{n-1}(t-(g_{n+1}-1)a/v) -u^y_n(t)) \cr+ &\theta(a/2v-t)(u^y_{n-1}(t-g_{n+1}a/v) -u^y_{n}(t))}-2u^y_n(t)]\cr
+{\sqrt 3\over   4}(\kr-\kp)
&[\eqalign{&u^x_{n+1}(t-(g_{n+1}-1)a/v)-u^x_{n+1}(t-g_{n+1}a/v)\cr -&\theta(t)(u^x_{n-1}(t-(g_{n+1}-1)a/v)-u^x_n(t))\cr +&\theta(a/2v-t) (
u^x_{n-1}(t-g_{n+1}a/v)-u^x_n(t))}] \cr
+\kr&[u^y_{n}(t+a/v)+u^y_{n}(t-a/v)-2u^y_{n}(t)]}\EQN TR12;b$$
Now take the Fourier transform in time of these equations. For the
layers with $n>1/2$ one has
$$F^x=\eqalign{ ({1\over 4}\kp+{3\over
4}\kr)&[\eqalign{&u^x_{n+1}(\omega)(e^{i\omega(g_{n+1}-1)a/v}+e^{i\omega g_{n+1}a/v})\cr
+&u^x_{n-1}(\omega)(e^{i\omega (g_{n+1}-1)a/v} + e^{(t-g_{n+1}a/v)})} -4u^x_{n}(\omega)]\cr
+{\sqrt 3\over   4}(\kr-\kp) &[\eqalign{&u^y_{n+1}(\omega)(e^{i\omega
(g_{n+1}-1)a/v}-e^{i\omega g_{n+1}a/v})\cr
-&u^y_{n-1}(\omega)(e^{i\omega (g_{n+1}-1)a/v} -e^{i\omega g_{n+1}a/v})}] \cr
+\kp&[u^x_{n}(\omega)e^{i\omega a/v}+e^{-i\omega a/v}-2]}\EQN TR12;a$$
$$F^y=\eqalign{ ({1\over 4}\kr+{3\over
4}\kp)&[\eqalign{&u^y_{n+1}(\omega)(e^{i\omega(g_{n+1}-1)a/v}+e^{i\omega g_{n+1}a/v})\cr
+&u^y_{n-1}(\omega)(e^{i\omega (g_{n+1}-1)a/v} + e^{(t-g_{n+1}a/v)})} -4u^y_{n}(\omega)]\cr
+{\sqrt 3\over   4}(\kr-\kp) &[\eqalign{&u^x_{n+1}(\omega)(e^{i\omega
(g_{n+1}-1)a/v}-e^{i\omega g_{n+1}a/v})\cr
-&u^x_{n-1}(\omega)(e^{i\omega (g_{n+1}-1)a/v} -e^{i\omega g_{n+1}a/v})}] \cr
+\kr&[u^y_{n}(\omega)e^{i\omega a/v}+e^{-i\omega a/v}-2]}\EQN TR12;b$$

Substituting in  the form
$$\pmatrix{u_n^x\cr u_n^y}=y^ne^{-i\omega
g_n/2v}\pmatrix{U_x\cr U_y}\EQN TR13$$
gives
$$-(m\omega^2+i\omega b(\omega)) U_x=\eqalign{\lb \eqalign{&(\kp+3\kr)\cos(\omega
a/2v){(y+y^{-1})\over 2} \cr + 2&\kp \cos(\omega a/v)-3(\kr+\kp)}\rb &U_x\cr
-\sqrt{3}i(\kr-\kp)\sin(\omega a/2v){(y-y^{-1})\over 2}& U_y}\EQN TR14;a$$
$$-(m\omega^2 +i\omega b(\omega))U_y=\eqalign{\lb \eqalign{&(\kr+3\kp)\cos(\omega
a/2v){(y+y^{-1})\over 2} \cr + 2&\kr \cos(\omega a/v)-3(\kr+\kp)}\rb &U_y\cr
-\sqrt{3}i(\kr-\kp)\sin(\omega a/2v){(y-y^{-1})\over 2}& U_x}.\EQN TR14;b$$
Here, and in what follows, the dissipation coefficient $b$ is
understood to be of the form
$$b(\omega)=b_0\sqrt{\omega^2+\omega^2_0}.\EQN TR13.1.1$$
The condition that the determinant of this system vanish determines
$y$ by 
$$0=\eqalign{&[\eqalign{&(\kp+3\kr)\cos(\omega
a/2v){(y+y^{-1})\over 2} \cr +&m\omega^2+i\omega b+2\kp \cos(\omega a/v)-3(\kr+\kp)}]
[\eqalign{&(\kr+3\kp)\cos(\omega
a/2v){(y+y^{-1})\over 2} \cr +&m\omega^2+i\omega b+2\kr \cos(\omega a/v)-3(\kr+\kp)}]\cr
&+3[(\kr-\kp)\sin(\omega a/2v){(y-y^{-1})\over 2}]^2}.\EQN TR14.1$$
It is more convenient to define
$$z={y+1/y\over 2}.\EQN TR14.2$$
Then the determinental condition becomes
$$\eqalign{&A=3(\kr-\kp)^2+16\kr\kp\cos(a\omega/(2v))^2\cr
&B=\cos(a\omega/2v)(2(3\kp^2+2\kp\kr+3\kr^2)\cos(a\omega/v)+4(\kr+\kp)
(m\omega^2+i\omega b-3(\kp+\kr)))\cr
&C=(m\omega^2+i\omega b-(\kp+\kr)(3-\cos(a\omega/v)))^2-
(\kp-\kr)^2(\cos(a\omega/v)^2+3\sin(a\omega/(2v))^2)\cr
&z_\pm={-B\pm\sqrt{B^2-4AC}\over 2A}}.\EQN TR14.3$$
There are now 4 values of $y$ which satisfy \Eq{TR14.1} for any given
$\omega$, namely
$$y_\pm=z_\pm+\sqrt{(z_\pm)^2-1}\EQN TR14.4$$
and two others given by the inverse of these, or equivalently by
changing the sign of the square root.

Define
$$\eqalign{&D_\pm= \eqalign{&m\omega^2 +i\omega b +(\kp+3\kr)\cos(\omega
a/2v){(y_\pm+y_\pm^{-1})\over 2} \cr + 2&\kp \cos(\omega a/v)-3(\kr+\kp)}\cr
&E_\pm=-\sqrt{3}i(\kr-\kp)\sin(\omega a/2v){(y^pm-y_\pm^{-1})\over
  2}}.\EQN TR14.5$$
Then a general solution of \Eq{TR12} is 
$$\pmatrix{u_n^x\cr u_n^y}=
\eqalign{&e^{-i\omega g_n/2v}[\eqalign{&y_+^{(n-1/2)}\pmatrix{E_+\cr-D_+}
u_{1+}+y_+^{(-n+1/2)}\pmatrix{E_+\cr D_+}u_{2+}\cr +
&y_-^{(n-1/2)}\pmatrix{E_-\cr-D_-}
u_{1-}+y_-^{(-n+1/2)}\pmatrix{E_-\cr D_-}u_{2-}}]\cr
&+U_N{(n-1/2)\over N}\pmatrix{0\cr 1}}.\EQN TR14.6$$

The four functions $u_{1\pm},u_{2\pm}$ can be determined from the four
conditions 
$$u^x_{N+1/2}=u^y_{N+1/2}-U_N=0\EQN TR14.7$$
and
$$\pmatrix{u_{1/2}^x\cr u_{1/2}^y}=
[\pmatrix{E_+\cr-D_+}
u_{1+}+\pmatrix{E_+\cr D_+}u_{2+} +
\pmatrix{E_-\cr-D_-}
u_{1-}+\pmatrix{E_-\cr D_-}u_{2-}].\EQN TR14.8$$
Once they are determined, one can obtain in particular a solution for
$u_{3/2}$, in terms of $u^x_{1/2}$ and $u^y_{1/2}$; however, the
expressions are too long to list here explicitly. 

Thus the problem is reduced to that of finding $u^x_{1/2}$ and
$u^y_{1/2}$. 
These are determined by taking the Fourier transform of the equations
on the line $n=1/2$. Unfortunately, we do not know how to solve the
equations in full generality. The one restriction that must be imposed
is that right on the crack line, we must take $\kr=0$. Otherwise the
formalism dies. We would like very much to overcome this restriction,
but do not now see how to do it.  However, to try to make up for it, we
will let $\kp$ equal some arbitrary $\kp^I$ on the interface. Given
this restriction, \Eq{TR12} becomes 
$$F^x=\eqalign{ &[\eqalign{&({1\over 4}\kp+{3\over
4}\kr)[u^x_{{3/2}}(t)+u^x_{3/2}(t-a/v)-2u^x_{1/2}(t)]\cr
+&{\kp^I\over 4}\theta(-t)(u^x_{-1/2}(t)-u_{1/2}^x(t))\cr +
&{\kp^I\over 4}\theta(a/2v-t)(u^x_{-1/2}(t-a/v)-u^x_1/2(t))} ]\cr
+&{\sqrt 3\over   4}
[\eqalign{&(\kr-\kp)[u^y_{3/2}(t)-u^y_{3/2}(t-a/v)]\cr +&
\kp^I\theta(-t)(u^y_{-1/2}(t)-u^y_{1/2}(t))\cr -
&\kp^I\theta(a/2v-t)(u^y_{-1/2}(t-a/v)-u^y_{1/2}(t))}] \cr
+\kp&[u^x_{1/2}(t+a/v)+u^x_{1/2}(t-a/v)-2u^x_{1/2}(t)]}\EQN TR15.5;a $$
$$F^y=\eqalign{ &[\eqalign{&({1\over 4}\kr+{3\over
4}\kp)[u^y_{{3/2}}(t)+u^y_{3/2}(t-a/v)-2u^y_{1/2}(t)]\cr
+&{3\kp^I\over 4}\theta(-t)(u^y_{-1/2}(t)-u_{1/2}^y(t))\cr +
&{3\kp^I\over 4}\theta(a/2v-t)(u^y_{-1/2}(t-a/v)-u^y_{1/2}(t))} ]\cr
+&{\sqrt 3\over   4}
[\eqalign{&(\kr-\kp)[u^x_{3/2}(t)-u^x_{3/2}(t-a/v)]\cr +&
\kp^I\theta(-t)(u^x_{-1/2}(t)-u^x_{1/2}(t))\cr -
&\kp\theta(a/2v-t)(u^x_{-1/2}(t-a/v)-u^x_{1/2}(t))}] \cr
+\kr&[u^y_{1/2}(t+a/v)+u^y_{1/2}(t-a/v)-2u^y_{1/2}(t)]}\EQN TR15.5;b$$

The important property of this set of equations is that there is
really only one linear combination of $u^x$ and $u^y$ which multiplies
the $\theta$ functions. This combination is
$$U(t)={-1\over 2\sqrt{3}}[u^x_{1/2}(t+a/2v)-u^x_{1/2}(t)]+{1\over 2} \lb
u^y_{1/2}(t+a/2v)+u^y_{1/2}(t)\rb\EQN TR16$$

To see why it enters, notice that when the strip is loaded in Mode I,
one must have the symmetries
$$u^y_{-1/2}(t)=-u^y_{1/2}(t+a/2v)\EQN TR17;a$$
$$u^x_{-1/2}(t)=u^x_{1/2}(t+a/2v).\EQN TR17;b$$
This symmetry allows one to eliminate the fields with subscript $-1/2$
from \Eq{TR15.5}.
The result is
$$F^x=\eqalign{ &({1\over 4}\kp+{3\over
4}\kr)[u^x_{{3/2}}(t)+u^x_{3/2}(t-a/v)-2u^x_{1/2}(t)]\cr
+&{\sqrt 3\over   4}
(\kr-\kp)[u^y_{3/2}(t)-u^y_{3/2}(t-a/v)]
\cr
+\kp&[u^x_{1/2}(t+a/v)+u^x_{1/2}(t-a/v)-2u^x_{1/2}(t)]
\cr -&{\kp^I\over 8\sqrt{3}}[U(t)\theta(-t)-U(t-a/2v)\theta(a/2v-t)]}\EQN
TR18.5;a $$
\medskip
$$F^y=\eqalign{ &({1\over 4}\kr+{3\over
4}\kp)[u^y_{{3/2}}(t)+u^y_{3/2}(t-a/v)-2u^y_{1/2}(t)]\cr
+&{\sqrt 3\over   4}
(\kr-\kp)[u^x_{3/2}(t)-u^x_{3/2}(t-a/v)]\cr 
+\kr&[u^y_{1/2}(t+a/v)+u^y_{1/2}(t-a/v)-2u^y_{1/2}(t)]
\cr -&{\kp^I\over 8}[U(t)\theta(-t)+U(t-a/2v)\theta(a/2v-t)]}\EQN TR18.5;b$$

It is now possible to Fourier transform \Eq{TR18.5}. The result is
$$-(m\omega^2 +ib\omega)u^x_{1/2}(\omega)=\eqalign{ &({1\over 4}\kp+{3\over
4}\kr)[u^x_{{3/2}}(\omega)(1+e^{i\omega a/v})-2u^x_{1/2}(\omega)]\cr
+&{\sqrt 3\over   4}
(\kr-\kp)[u^y_{3/2}(\omega)(1-e^{i\omega a/v})]
\cr
+\kp&[2u^x_{1/2}\cos (a\omega/v)-2u^x_{1/2}(\omega)]
\cr -&{\kp^I\over 8\sqrt{3}}U^-(\omega)(1-e^{i\omega a/2v})}\EQN TR18.5;a $$
\medskip
$$-(m\omega^2+ib\omega) u^y_{1/2}(\omega)=\eqalign{ &({1\over 4}\kr+{3\over
4}\kp)[u^y_{{3/2}}(\omega)(1+e^{i\omega a/v})-2u^y_{1/2}(\omega)]\cr
+&{\sqrt 3\over   4}
(\kr-\kp)[u^x_{3/2}(\omega)(1-e^{i\omega a/v})]\cr 
+\kr&[2 u^y_{1/2}(\omega)\cos(a\omega/v)-2u^y_{1/2}(\omega)]
\cr -&{\kp^I\over 8}U^-(\omega)(1+e^{i\omega a/v})}  \EQN TR18.5;b$$

Using \Eq{TR14.6} to find $u^x_{3/2}$ and$u^y_{3/2}$, 
and \Eq{TR16} to find  $U(\omega)$, one can eliminate all variables
but $U$ from \Eq{TR18.5}. 

Analyzing the  $\omega\rightarrow 0$ behavior, where
$$\eqalign{ &u^x_{1/2}\rightarrow 0\cr
&U\rightarrow u^y_{1/2},\cr
& u^y_{3/2}\rightarrow (1-1/N)u^y_{1/2}+U_N/N\delta(\omega)}\EQN TR19$$
one finds finally that
$$Q(\omega)U^++U^-=Q_0U_N\delta(\omega)
   \EQN TR20$$
as in Section \use{sec.1D}. The expression for $Q$ is again too
lengthy to record here.

\end